\documentclass[12pt]{article}
\usepackage{graphicx,epsfig,amsmath}

\def\beq{\begin{equation}}
\def\eeq{\end{equation}}
\def\bea{\begin{eqnarray}}
\def\eea{\end{eqnarray}}
\def\bmt{\begin{pmatrix}}
\def\emt{\end{pmatrix}}
\def\ba{\begin{align*}}
\def\ea{\end{align*}}
\def\bal{\begin{align}}
\def\eal{\end{align}}
\def\bg{\begin{gather*}}
\def\eg{\end{gather*}}
\def\bga{\begin{gather}}
\def\ega{\end{gather}}

\def\a{\alpha}
\def\ep{\epsilon}

\def\g{\gamma}
\def\sla#1{/\kern-.6em#1}
\def\sbreak{\kern.7em/\kern-1emSUSY}

\begin{document}
\thispagestyle{empty} \noindent
\begin{flushright}
        OHSTPY-HEP-T-06-004     \\
        June 2006
\end{flushright}

\vspace{1cm}
\begin{center}
  \begin{large}
  \begin{bf}

SO(10) SUSY GUT for Fermion Masses :

Lepton Flavor and CP Violation

\end{bf}
 \end{large}
\end{center}
  \vspace{1cm}
     \begin{center}
R. Derm\' \i \v sek$^\dagger$, M. Harada$^*$ and S. Raby$^*$\\
      \vspace{0.3cm}
\begin{it}
$^\dagger$School of Natural Sciences, Institute for Advanced Study,  Princeton, NJ 08540 \\
$^*$Department of Physics, The Ohio State University, 191 W.
Woodruff Ave., Columbus, Ohio  43210
\end{it}
  \end{center}
  \vspace{1cm}
\centerline{\bf Abstract}
\begin{quotation}
\noindent We discuss the results of a global $\chi^2$ analysis of a
simple $SO(10)$ SUSY GUT with $D_3$ family symmetry and low energy R
parity. The model describes fermion mass matrices with 14 parameters
and gives excellent fits to 20 observable masses and mixing angles
in both quark and lepton sectors, giving 6 predictions. Bi-large
neutrino mixing is obtained with hierarchical quark and lepton
Yukawa matrices; thus avoiding the possibility of large lepton
flavor violation. The model naturally predicts small 1-3 neutrino
mixing,  with $\sin \theta_{13} \simeq 0.05 - 0.06$.   In this paper
we evaluate the predictions for the lepton flavor violating
processes, $\mu \rightarrow e \gamma$, $\tau \rightarrow  \mu
\gamma$ and $\tau \rightarrow  e \gamma$ and also the electric
dipole moment of the electron, $d_e$, muon and tau, assuming
universal squark and slepton masses, $m_{16}$, and a universal soft
SUSY breaking A parameter, $A_0$, at the GUT scale. We find $Br(\mu
\rightarrow e \gamma)$ is naturally below present bounds, but may be
observable by MEG. Similarly, $d_e$ is below present bounds; but is
within the range of future experiments. We also give predictions for
the light Higgs mass (using FeynHiggs). We find an upper bound given
by $m_h \leq 127$ GeV, with an estimated $\pm 3$ GeV theoretical
uncertainty. Finally we present predictions for SUSY particle masses
in the favored region of parameter space.

\end{quotation}

\newpage

\section{Introduction}

In this letter we present results for a global $\chi^2$ analysis of
the $SO(10)$ SUSY GUT for fermion masses presented in
Ref.~\cite{Dermisek:2005ij}.  The model also has a $D_3 \times [U(1)
\times Z_2 \times Z_3]$ family symmetry.\footnote{The charged
fermion sector of this theory was considered in an earlier paper
\cite{Dermisek:1999vy} and the neutrino sector of the theory was
inspired by the previous analysis by one of us, R.D.
\cite{Dermisek:2004tx}.}   The three families of quarks and leptons
are contained in three 16 dimensional representations of $SO(10)$ $
\{ 16_a, \; 16_3 \}$ with $16_a, \; a = 1,2$ a $D_3$ flavor doublet
(see Ref.~\cite{Dermisek:1999vy} for details on $D_3$).  The third
family, along with the pair of electroweak Higgs doublets, contained
in the $10$ dimensional representation of $SO(10)$, are $D_3$
singlets.   Hence only the third generation has a renormalizable
Yukawa coupling and, as a consequence, we have $\lambda_t =
\lambda_b = \lambda_\tau = \lambda_{\nu_\tau}$ Yukawa unification at
$M_{GUT}$. This forces us into the large $\tan\beta$ regime and
several interesting predictions follow.   We have derived the
consequences of third generation Yukawa unification in several
papers.  In Ref. \cite{bdr} we demonstrated that in order to fit the
low energy values of the top, bottom and tau masses (with the
typically large, of order 50\%, radiative corrections to the bottom
quark mass) the soft SUSY breaking parameters necessarily reside in
a very narrow region of the possible parameter space.   Hence we
have definite predictions for SUSY spectra, see \cite{bdr} and
Section \ref{sec:results} for more details.  In addition, in this
region of parameter space the light Higgs mass necessarily has a
central value of order 120 GeV.   In Ref. \cite{Dermisek:2003vn} we
demonstrated that this same minimal $SO(10)$ SUSY model
[MSO$_{10}$SM] gives the correct abundance of dark matter, fitting
the WMAP data,  and gives observable values for the branching ratio
$Br(B_s \rightarrow \mu^+ \ \mu^-)$ with a lower bound of order
$10^{-8}$.   The dark matter candidate in this model is the LSP,
neutralino, which predominantly annihilates through a direct
s-channel CP odd Higgs, $A$.   In addition, it also dominates in the
leptonic decay of $B_s$.

In the present model,  all of the above results are retained (with
small modifications), but in addition we fit the masses and mixing
angles of all three families, including neutrino data. The model
describes fermion mass matrices with 14 parameters and gives
excellent fits to 20 observable masses and mixing angles in both
quark and lepton sectors, giving 6 predictions.  Both the charged
fermion and neutrino mass matrices are hierarchical; thus
suppressing large flavor violating interactions, even at large
$\tan\beta$.   The simple structure of the neutrino sector leads
quite naturally to maximal atmospheric neutrino oscillations and
large solar neutrino mixing \cite{Dermisek:2005ij}. We predict a
very small value for $\sin\theta_{13} \simeq 0.05 - 0.06$.   In
addition, CP violation in the neutrino sector is fixed by the phases
in the charged fermion mass matrices. At the same we can easily
accommodate leptogenesis via non-thermal processes, see for example
\cite{Senoguz:2003hc}.

\section{The Model \label{sec:model}}
The full superpotential $W = W_{ch. fermions} + W_{neutrino}$ for
fermion masses and mixing angles contains two terms. The first term,
resulting in Dirac Yukawa matrices for charged fermions and
neutrinos, is given by
\begin{eqnarray} W_{ch. fermions} = & 16_3 \ 10 \ 16_3 +  16_a \ 10 \ \chi_a & \label{eq:chfermionD3}
\\ & +  \bar \chi_a \ ( M_{\chi} \ \chi_a + \ 45 \ \frac{\phi_a}{\hat M} \ 16_3 \ + \ 45 \ \frac{\tilde
\phi_a}{\hat M} \  16_a + {\bf A} \ 16_a ). & \nonumber
\end{eqnarray}
The third family of quarks and leptons is contained in the
superfield  $16_3$ (transforming as a $16$ of $SO(10)$); the first
two families are contained in $16_a, \ a = 1,2$ (with explicit
$SO(10)$ tranformations and transforming as a $D_3$ doublet) and the
two Higgs doublets are contained in $10$.   The additional fields
are an adjoint of $SO(10)$ ($45$) and several $SO(10)$ singlet
flavon fields needed to break the full flavor symmetry $D_3 \times
[U(1) \times Z_2 \times Z_3]$. Note, $M_{\chi} = M_0 ( 1 + \alpha X
+ \beta Y )$ includes $SO(10)$ breaking vevs in the $X$ and $Y$
directions, $\phi^a, \ \tilde \phi^a \, (D_3 \; {\rm doublets}), \
{\bf A} \; ({\bf 1_B} \; {\rm singlet })$ are $SO(10)$ singlet
flavon fields, and $\hat M, \ M_0$ are $SO(10)$ singlet masses.  The
fields $45, \; {\bf A}, \ \phi, \ \tilde \phi \;$ are assumed to
obtain vevs $\langle 45 \rangle \sim (B - L) \ M_G$ (where $B, \ L,
\ M_G$ is baryon and lepton number and the GUT scale, respectively)
, $\; {\bf A} \ll M_0 \;$ and
\begin{equation} \langle \phi \rangle =  \left( \begin{array}{c} \phi_1 \\
\phi_2 \end{array} \right), \; \langle \tilde \phi \rangle = \left(
\begin{array}{c} 0 \\ \tilde \phi_2
\end{array} \right) \end{equation} with $\phi_1 > \phi_2$.  The second term gives large
lepton number violating masses for ``right-handed" neutrinos;
necessary for the See-Saw mechanism.  We have
\begin{eqnarray} \label{eq:WneutrinoD3} W_{neutrino} = &
\overline{16} \left(\lambda_2 \ N_a \ 16_a \ + \ \lambda_3 \ N_3 \ 16_3 \right) & \\
& +  \;\; \frac{1}{2} \left(S_{a} \ N_a \ N_a \;\; + \;\; S_3 \ N_3
\ N_3\right)   & \nonumber
\end{eqnarray}
where the fields $N_3; \ N_a \ a = 1,2$ are $SO(10)$ singlets and
$\overline{16}$ is assumed to break $SO(10)$ to $SU(5)$ via a vev in
the right-handed neutrino direction.

The superpotential, (Eqn. \ref{eq:chfermionD3}) results in the
following Yukawa matrices:\footnote{In our notation, Yukawa matrices
couple electroweak doublets on the left to singlets on the right. It
has been shown in Ref. \cite{Blazek:1999ue} that excellent fits to
charged fermion masses and mixing angles are obtained with this
Yukawa structure.}
\begin{eqnarray}
Y_u =&  \left(\begin{array}{ccc}  0  & \epsilon' \ \rho & - \epsilon \ \xi  \\
             - \epsilon' \ \rho &  \tilde \epsilon \ \rho & - \epsilon     \\
       \epsilon \ \xi   & \epsilon & 1 \end{array} \right) \; \lambda & \nonumber \\
Y_d =&  \left(\begin{array}{ccc}  0 & \epsilon'  & - \epsilon \ \xi \ \sigma \\
- \epsilon'   &  \tilde \epsilon  & - \epsilon \ \sigma \\
\epsilon \ \xi  & \epsilon & 1 \end{array} \right) \; \lambda & \label{eq:yukawaD3} \\
Y_e =&  \left(\begin{array}{ccc}  0  & - \epsilon'  & 3 \ \epsilon \ \xi \\
          \epsilon'  &  3 \ \tilde \epsilon  & 3 \ \epsilon  \\
 - 3 \ \epsilon \ \xi \ \sigma  & - 3 \ \epsilon \ \sigma & 1 \end{array} \right) \; \lambda &
 \nonumber \\
Y_{\nu} =&  \left(\begin{array}{ccc}  0  & - \epsilon' \ \omega & {3 \over 2} \ \epsilon \ \xi \ \omega \\
      \epsilon'  \ \omega &  3 \ \tilde \epsilon \  \omega & {3 \over 2} \ \epsilon \ \omega \\
       - 3 \ \epsilon \ \xi \ \sigma   & - 3 \ \epsilon \ \sigma & 1 \end{array} \right) \; \lambda &
 \end{eqnarray}
with  \begin{eqnarray}  \xi \;\; =  \;\; \phi_2/\phi_1; & \;\;
\tilde \epsilon  \;\; \propto   \;\; \tilde \phi_2/\hat M;  & \label{eq:omegaD3} \\
\epsilon \;\; \propto  \;\; \phi_1/\hat M; &  \;\;
\epsilon^\prime \;\; \sim  \;\;  ({\bf A}/M_0); \nonumber \\
  \sigma \;\; =   \;\; \frac{1+\alpha}{1-3\alpha}; &  \;\; \rho \;\; \sim   \;\;
  \beta \ll \alpha & \nonumber \\
 & \omega \;\; =  \;\; 2 \, \sigma/( 2 \, \sigma - 1) .  & \nonumber \end{eqnarray}  The Dirac
mass matrices are then given by
 \begin{eqnarray} m_i \equiv Y_i \frac{v}{\sqrt{2}} \sin\beta  &   i = \nu, \ u & \\
  m_i \equiv Y_i \frac{v}{\sqrt{2}} \cos\beta  &   i = e, \ d &  .
 \label{eq:mnuD3}
  \end{eqnarray}

Consider the neutrino masses.  In the three $16$s we have three
electroweak doublet neutrinos ($\nu_a, \ \nu_3$) and three
electroweak singlet anti-neutrinos ($\bar \nu_a, \ \bar \nu_3$). In
addition, the anti-neutrinos get GUT scale masses by mixing with
three $SO(10)$ singlets $\{ N_a, \ a = 1,2; \;\; N_3 \}$
transforming as a $D_3$ doublet and singlet respectively. We assume
$\overline{16}$ obtains a vev, $v_{16}$, in the right-handed
neutrino direction, and $\langle S_{a} \rangle = M_a$ for $a = 1,2$
(with $M_2 > M_1$) and $\langle S_3 \rangle = M_3$.\footnote{These
are the most general set of vevs for $\phi_a$ and  $S_{a}$.  The
zero vev for $\tilde \phi_1$ can be enforced with a simple
superpotential term such as $S \ \tilde \phi_a \ \tilde \phi_a$.  }
We thus obtain the effective neutrino mass terms given by
\begin{equation} W =  \nu \ m_\nu \ \bar \nu + \bar \nu \ V \ N +
\frac{1}{2} \ N \ M_N \ N \end{equation} with
\begin{equation} V = v_{16} \ \left(
\begin{array}{ccc} 0 &  \lambda_2 & 0 \\
\lambda_2 & 0 & 0 \\ 0 & 0 &  \lambda_3 \end{array} \right), \; M_N
= diag( M_1,\ M_2,\ M_3) . \end{equation}

The electroweak singlet neutrinos $ \{ \bar \nu, N \} $ have large
masses $V , M_N \sim M_G$.  After integrating out these heavy
neutrinos, we obtain the light neutrino mass matrix given by
\begin{equation}
{\cal M} =   m_\nu  \ M_R^{-1} \ m_\nu^T ,
\end{equation}
where the effective right-handed neutrino Majorana mass matrix is
given by:
\begin{equation}
M_R =  V \ M_N^{-1}  \ V^T  \  \equiv \  {\rm diag} ( M_{R_1},
M_{R_2}, M_{R_3} ),
\end{equation}
with \bea M_{R_1} = (\lambda_2 \ v_{16})^2/M_2, \quad  M_{R_2} =
(\lambda_2 \ v_{16})^2/M_1, \quad  M_{R_3} = (\lambda_3 \
v_{16})^2/M_3 . \label{eq:rhmass} \eea  Defining $U_e$ as the
$3\times3$ unitary matrix for left-handed leptons needed to
diagonalize $Y_e$ (Eqn. \ref{eq:yukawaD3}), i.e. $Y_e^D = U_e^T \
Y_e \ U_{\bar e}^*$ and also $U_\nu$ such that $U_\nu^T \ {\cal M} \
U_\nu = {\cal M}_D = {\rm diag}( m_{\nu_1}, \ m_{\nu_2}, \
m_{\nu_3})$, then the neutrino mixing matrix is given by $U_{PMNS} =
U_e^\dagger \ U_\nu$ in terms of the flavor eigenstate
($\nu_\alpha$, $\alpha =  e, \ \mu, \ \tau$) and mass eigenstate
($\nu_i$, $i = 1,2,3$) basis fields with \bea \nu_\alpha = \sum_i
(U_{PMNS})_{\alpha i} \ \nu_i . \eea

For $U_{PMNS}$ we use the notation of Ref~\cite{Eidelman:2004wy} with  \bea \left( \begin{array}{c} \nu_e \\
\nu_\mu \\ \nu_\tau \end{array} \right) = \left(
\begin{array}{ccc} c_{1 2} c_{1 3} & s_{1 2} c_{1 3} & s_{1 3} e^{-i\delta} \\ -s_{12} c_{23} - c_{12} s_{23}
s_{13} e^{i\delta} & c_{12} c_{23} - s_{12} s_{23} s_{13}
e^{i\delta} & s_{23} c_{13} \\ s_{12} s_{23} - c_{12} c_{23} s_{13}
e^{i\delta} & -c_{12} s_{23} - s_{12} c_{23} s_{13} e^{i\delta} &
c_{23} c_{13} \end{array} \right) \left(
\begin{array}{c} e^{i\alpha_1 /2} \nu_1 \\ e^{i\alpha_2 /2}  \nu_2 \\ \nu_3 \end{array} \right) \label{eq:PMNS}
\eea

Finally, we note that this theory is certainly not fundamental with
many arbitrary symmetry breaking VEVs at the GUT scale.
Nevertheless, it has two major features in its favor.   As a result
of the GUT and family symmetries, the Yukawa matrices, which are the
only observables of the complicated GUT physics, have fewer
parameters than low energy observables.  Hence this theory is
predictive. Secondly, the model has the advantage that it
self-consistently fits the low energy data, and thus, at the very
least, it is an excellent phenomenological ansatz for fermion
masses.  Thus it can be tested via additional low energy flavor
violating processes.

\section{Global $\chi^2$ Analysis \label{sec:chi2}}

Yukawa matrices in this model are described by seven real parameters
\{$\lambda$,  $\epsilon$, $\tilde \epsilon$, $\sigma$, $\rho$,
$\epsilon'$, $\xi $\} and, in general, four phases \{$\Phi_\sigma, \
\Phi_{\tilde \epsilon}, \ \Phi_\rho, \ \Phi_\xi $\}. Therefore, in
the charged fermion sector we have 11 parameters to explain 9 masses
and three mixing angles and one CP violating phase in the CKM
matrix, leaving us with 2 predictions.\footnote{Of course, in any
supersymmetric theory there is one additional parameter in the
fermion mass matrices, i.e. $\tan\beta$.  Including this parameter,
there is one less prediction for fermion masses, but then (once SUSY
is discovered) we have one more prediction.  This is why we have not
included it explicitly in the preceding discussion. } Note, these
parameters also determine the neutrino Yukawa matrix. Finally, our
minimal ansatz for the right-handed neutrino mass matrix is given in
terms of three additional real parameters\footnote{In principle,
these parameters can be complex. We will nevertheless assume that
they are real; hence there are no additional CP violating phases in
the neutrino sector.}, i.e. the three right-handed neutrino masses.
At this point the three light neutrino masses and the neutrino
mixing matrix, $U_{PMNS}$, (3+4 observables) are completely
specified. Altogether, the model describes 20 observables in the
quark and lepton sectors with 14 parameters, effectively having 6
predictions.\footnote{Note, the two Majorana phases are in principle
observable, for example, in neutrinoless double-beta decay
\cite{Petcov:2004wz}, however, the measurement would be very
difficult (perhaps too difficult \cite{Barger:2002vy}).   If
observable, this would increase the number of predictions to 8.}

In addition to the parameters describing the fermion mass matrices,
we have to input three parameters specifying the three gauge
couplings: the GUT scale $M_G$ defined as the scale at which
$\alpha_1$ and $\alpha_2$ unify, the gauge coupling at the GUT scale
$\alpha_G$, and the correction $\epsilon_3$ to $\alpha_3(M_G)$
necessary to fit the low energy value of the strong coupling
constant. Finally we have to input 7  supersymmetry parameters given
by -- $M_{1/2}$, a universal gaugino mass; $m_{10}$, a universal
Higgs mass; $m_{16}$, a universal squark and slepton mass; $A_0$, a
universal trilinear coupling; a small Higgs mass splitting
parameter, $\Delta m_H = 1/2(m_{H_d}^2 - m_{H_u}^2)/m_{10}^2$;  the
supersymmetric Higgs mass parameter $\mu (M_Z)$ and the ratio of the
two Higgs VEVs,  $\tan\beta$.

We have also imposed some physically motivated constraints on the
$\chi^2$ analysis. We demand a lower bound on the lightest stop mass
given by $m_{\tilde t} = 500$ GeV. Lower values of $m_{\tilde t}$
actually give even better fits. On the other hand, a chargino-stop
loop gives the dominant SUSY contribution to the process $b
\rightarrow s \gamma$ and lighter stop values make it difficult to
fit this process. In addition we fix the CP odd Higgs mass, $m_A$ =
700 GeV. Lower values would result in a branching ratio $Br(B_s
\rightarrow \mu^+ \mu^-)$ which approaches the experimental lower
bound, see Ref. \cite{Dermisek:2003vn}.  Our results however are not
sensitive to this latter constraint. Further discussion of the
former constraint is given in Sect. \ref{sec:results}.

All the parameters (except for $\mu (M_Z)$) are run from the GUT
scale to the weak scale ($M_Z$) using two (one) loop RGEs for
dimensionless (dimensionful) parameters. At the weak scale, the SUSY
partners are integrated out leaving the two Higgs doublet model as
an effective theory. We require proper electroweak symmetry
breaking. Moreover, the full set of  one loop, electroweak and SUSY,
threshold corrections to fermion mass matrices, as well as to the
three precision electroweak observables ($G_{\mu}, \
\alpha_{EM}^{-1}$ and $\alpha_s(M_Z)$) are calculated at
$M_Z$.\footnote{In a top down analysis, the $\overline{DR}$ value of
$sin^2\theta_W(M_Z)$ is obtained directly via RG running
$\alpha_{(1,2)}$ from the GUT scale.  Then with the calculated value
of $M_Z$ and all one loop threshold corrections included in $\Delta
r$, we obtain the observed value of $G_\mu$ \cite{Pierce:1996zz}.}
Below $M_Z$ we use 3 loop QCD and 1 loop QED RG equations to
calculate light fermion masses and $\alpha_{EM}$. More details about
the analysis can be found in~\cite{bdr} or
\cite{Dermisek:2003vn}.\footnote{The only difference is that in the
present analysis we include all three families of fermions.}  In
addition, we self-consistently include the contributions of the
right-handed neutrinos to the RG running between the GUT scale and
the mass of the heaviest right-handed neutrino~\cite{neutrino-RGE}.

The $\chi^2$ function is constructed from observables given in
Table~\ref{t:fit4tev}.  We have used the top quark mass from the PDG
reviews of particle properties 2005 update by T. M. Liss and A.
Quadt.\footnote{This value of the top quark mass ($172.7 \pm 2.9$
Gev/c$^2$) is not so different from the most recent value ($172.5
\pm 1.3_{stat} \pm 1.9_{syst} {\textrm GeV}/c^2$) from CDF II and
DZero at the Tevatron \cite{Whiteson:2006zp}. } Note that we have
included several redundant observables in the quark sector. We do
this because quark masses are not known with high accuracy and
different combinations of quark masses usually have independent
experimental and theoretical uncertainties. Thus we include three
observables for the charm and bottom quark masses: the
$\overline{MS}$ running masses ($m_c(m_c), \ m_b(m_b)$) and the
difference in pole masses $M_b - M_c$ obtained from heavy quark
effective theory. The same is true for observables in the CKM
matrix.  For example, we include $V_{td}$ and the two CP violating
observables $\epsilon_K$ and the value for $\sin(2\beta)$ given by
the world average measured via the process $B \rightarrow J/\psi \
K_s$ \cite{Barberio:2006bi}. However we have doubled the error to
take into account the significant difference between the BaBar and
Belle central values. We thus have 16 observables in the quark and
charged lepton sectors. We use the central experimental values and
one sigma error bars from the particle data
group~\cite{Eidelman:2004wy}. However, in the case that the
experimental error is less than $0.1 \%$ we use $\sigma = 0.1 \%$
due to the numerical precision of our calculation.

At present only four observables in the neutrino sector are
measured. These are the two neutrino mass squared differences,
$\Delta m^2_{31} $ and $\Delta m^2_{21}$, and two mixing angles,
$\sin^2 \theta_{12}$ and $\sin^2 \theta_{23}$. For these observables
we use the central values and $2 \sigma$ errors from
Ref.~\cite{Maltoni:2004ei}. The other observables: neutrino masses,
1-3 mixing angle and the phase of the lepton mixing matrix are
predictions of the model.   In addition, the new feature of this
paper is the predictions for several lepton flavor violating
processes and lepton electric dipole moments.

\section{Results
\label{sec:results}}

Let us now discuss our results.  We performed the global $\chi^2$
analysis for values of the soft SUSY breaking scalar mass at $M_G$
given by $m_{16} =$ 3, 4 and 5 TeV.  However, we only present the
results for the latter two cases. Good fits prefer the region of
SUSY parameter space characterized by\footnote{In addition, the best
fit requires a non-universal Higgs masses at the GUT scale with
$\Delta m_H = 1/2(m_{H_d}^2 - m_{H_u}^2)/m_{10}^2 \sim .07$.  Note
this is significantly smaller than needed in the past~\cite{bdr}.
That is because the RGE running of neutrino Yukawas from $M_G$ to
the heaviest right handed neutrino has been included
self-consistently. As noted in \cite{bdr}, such running was a
possible source for Higgs splitting.  Evidently it can not be the
only source.}  \begin{eqnarray} \mu, \ M_{1/2} \ \ll \ m_{16}  \label{eq:susypara} \\
- A_0 \ \simeq \  \sqrt{2} m_{10} \ \simeq \ 2 m_{16}. \nonumber
\end{eqnarray} This is required in order to fit the top, bottom and
tau masses when the third generation Yukawa couplings
unify~\cite{bdr}.  Note, the three input parameters ($\mu, \
M_{1/2}, \ m_{16}$) are not varied when minimizing $\chi^2$.
As a consequence of the relations, (Eqn. \ref{eq:susypara}), we
expect heavy first and second generation squarks and sleptons, while
the third generation scalars are significantly lighter (with the
stop generically the lightest). In addition, charginos and
neutralinos are typically the lightest superpartners.  We predict
values of $\tan\beta \sim 50$ and a light Higgs with mass, $m_h \leq
127$ GeV (with a theoretical uncertainty $\pm 3$ GeV). The specific
relations between the SUSY breaking parameters also lead to an
interesting prediction for the process $B_s \rightarrow \mu^+ \mu^-
$ with branching ratio in the region currently being explored at the
Tevatron.\footnote{This process is sensitive to the CP odd Higgs
mass, $m_A$, which can be adjusted in theories with non-universal
Higgs masses.} Furthermore, the neutralino relic density obtained
for our best fit parameters is consistent with WMAP
data~\cite{Dermisek:2003vn} and direct neutralino detection is
possible in near future experiments. Finally, this region maximally
suppresses the dimension five contribution to proton
decay~\cite{Dermisek:2000hr} and suppresses SUSY flavor and CP
violation in general. For more information on the SUSY and Higgs
spectra and related phenomenology in this region of SUSY breaking
parameter space, see Refs.~\cite{bdr} and \cite{Dermisek:2003vn}.

\begin{figure}[t!]
\begin{center}
\begin{minipage}{6in}
\hspace*{5mm} \epsfig{figure=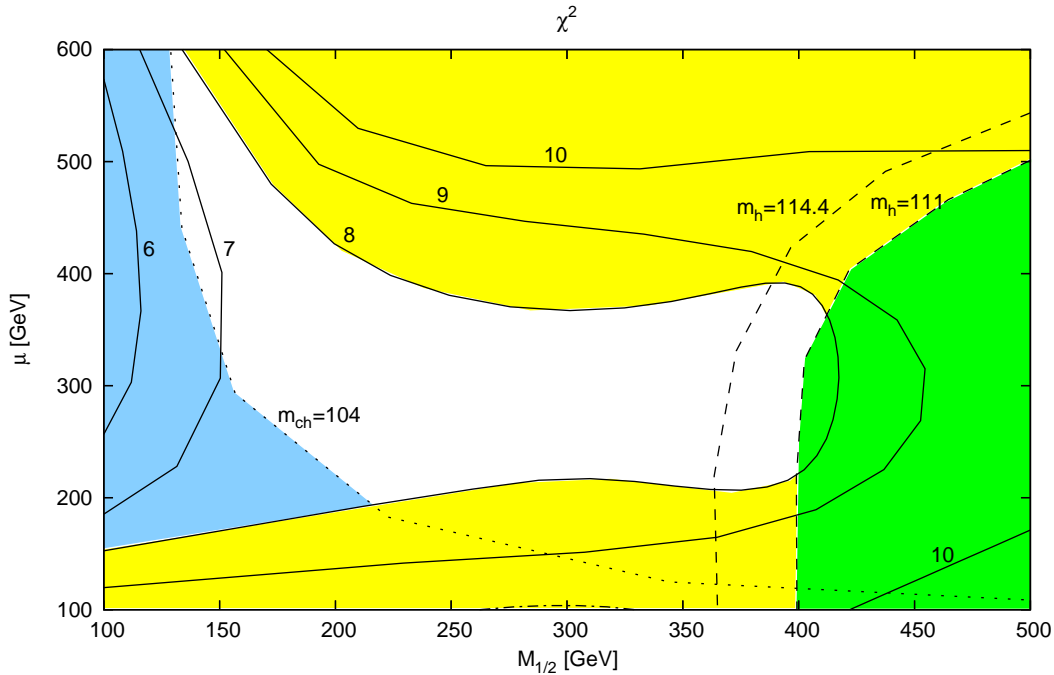,width=14cm}
\end{minipage}
\end{center}
\begin{center}
\caption{\label{fig:4TeVchi2} {\small Contours of constant $\chi^2$
for
    $m_{16} = 4$ TeV and $m_{A} = 700$ GeV.   The yellow (very light) shaded region at the bottom and top (and the region bounded by
    the extended solid boundary line) has $\chi^2 \geq 8$.  The blue (light shaded) region on the left (and below the
    extended dotted line) is excluded by $m_{\chi^+}<104$ GeV and the green (darker shaded) region is excluded
    by the Higgs mass bound $m_{h}<111$ GeV.  } }
\end{center}
\end{figure}
\begin{figure}[t!]
\begin{center}
\begin{minipage}{6in}
\hspace*{5mm} \epsfig{figure=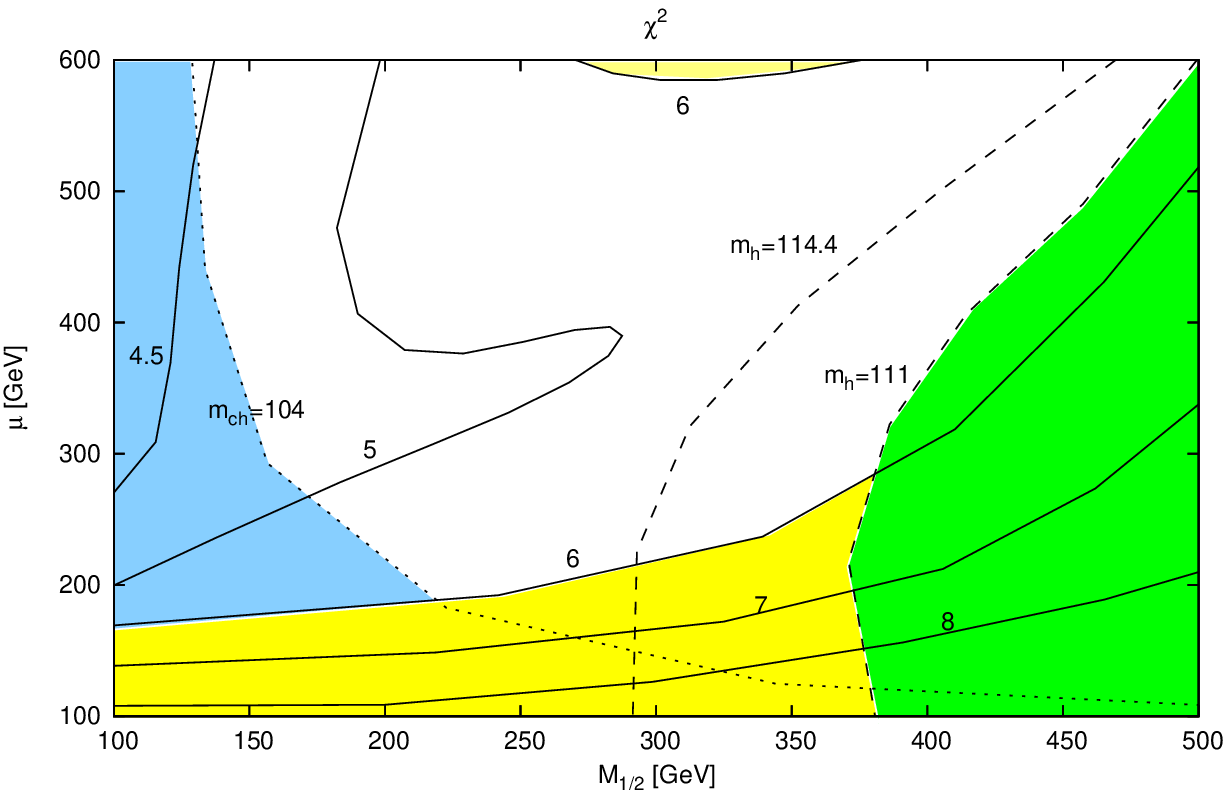,width=14cm}
\end{minipage}
\end{center}
\begin{center}
\caption{\label{fig:5TeVchi2} {\small Contours of constant $\chi^2$
for
    $m_{16} = 5$ TeV and $m_{A} = 700$ GeV.   The yellow (very light) shaded region at the bottom and top (and the region bounded by
    the extended boundary line) has $\chi^2 \geq 6$.  The blue (light shaded) region on the left (and below the
    extended dotted line) is excluded by $m_{\chi^+}<104$ GeV and the green (darker shaded) region is excluded
    by the Higgs mass bound $m_{h}<111$ GeV.  } }
\end{center}
\end{figure}
In Figs. \ref{fig:4TeVchi2} and \ref{fig:5TeVchi2} we present
contours of constant $\chi^2$ for $m_{16} = 4$ and $5$ TeV with
$m_{A} = 700$ GeV, as a function of $\mu, \ M_{1/2}$.  The best fits
are obtained for small values of $M_{1/2} \leq 300$ GeV (where the
lower bound on $M_{1/2}$ is determined by the experimental bound on
the chargino mass, $m_{\chi^+} > 104$ GeV). We find that the value
of $\chi^2$ decreases as $m_{16}$ increases. This is solely due to
the lower bound of 500 GeV on the lightest stop mass and the
resultant difficulty in fitting the bottom quark mass with heavier
stop.

At this point a brief aside is necessary.  In our analysis, we have
not evaluated several significant pieces of data.  These include the
branching ratios $Br(b \rightarrow s \gamma)$,  $Br(B \rightarrow
X_s \ l^+ \ l^-)$, and $B_s - \bar B_s$ mixing.   These also provide
significant constraints on the theory. However, we have used the
code of T. Bla\v{z}ek \cite{Blazek:1996yv, Blazek:1999ue} to check
our analysis and also evaluate the branching ratio $Br(b \rightarrow
s \gamma)$. This process is enhanced at large $\tan\beta$.   The
dominant SUSY contribution comes from the chargino-stop loop.   We
find that the most significant constraint is a lower bound on the
lightest stop mass of order 500 GeV.  We have thus imposed this
bound on the stop mass by introducing a penalty to $\chi^2$.  The
best fit for the branching ratio $Br(b \rightarrow s \gamma)$ is
then fit with the minimal stop mass. Note, we find that $\chi^2$
increases as the lower bound on the stop mass increases. This is due
to the fact that a good fit for $m_b$ prefers a light stop
\cite{bdr,Dermisek:2003vn}. In addition,  as the lower bound on the
stop mass increases,  we find it necessary to increase $m_{16}$. For
example, with the light stop mass, $m_{\tilde t} = 300$ GeV, we find
good fits with $m_{16} = 3$ TeV \cite{bdr}. Now with $m_{\tilde t} =
500$ GeV, good fits, with $\chi^2 \leq 8$, are only obtained with
$m_{16} \geq 4$ TeV.  As a final note, for a light stop ($m_{\tilde
t} = 500$ GeV) the Wilson coefficient of the dominant operator for
the process $b \rightarrow s \gamma$,  ${\cal O}_7$,  has the
opposite sign in the MSSM than for the standard model, i.e.
$C_7(MSSM) \sim - C_7(SM)$ \cite{Blazek:1995nv}. Recent measurements
of the branching ratio $Br(B \rightarrow X_s \ l^+ \ l^-)$
\cite{Gambino:2004mv} suggest that the same sign is preferred,
$C_7(MSSM) \sim C_7(SM)$; while the latest data on the
forward-backward asymmetry does not seem to distinguish these two
possibilities \cite{Ishikawa:2006fh}.  Clearly a more detailed
$\chi^2$ analysis including all these processes would be necessary
to better test the theory. This is however not the focus of the
present paper.\footnote{We have also not calculated $(g - 2)_\mu$ in
this paper.   However in a previous analysis \cite{Dermisek:2003vn}
we found typical values of $(g - 2)_\mu < 3 \times 10^{-10}$.}

\begin{figure}[t!]
\begin{center}
\begin{minipage}{6in}
\hspace*{5mm}
\epsfig{figure=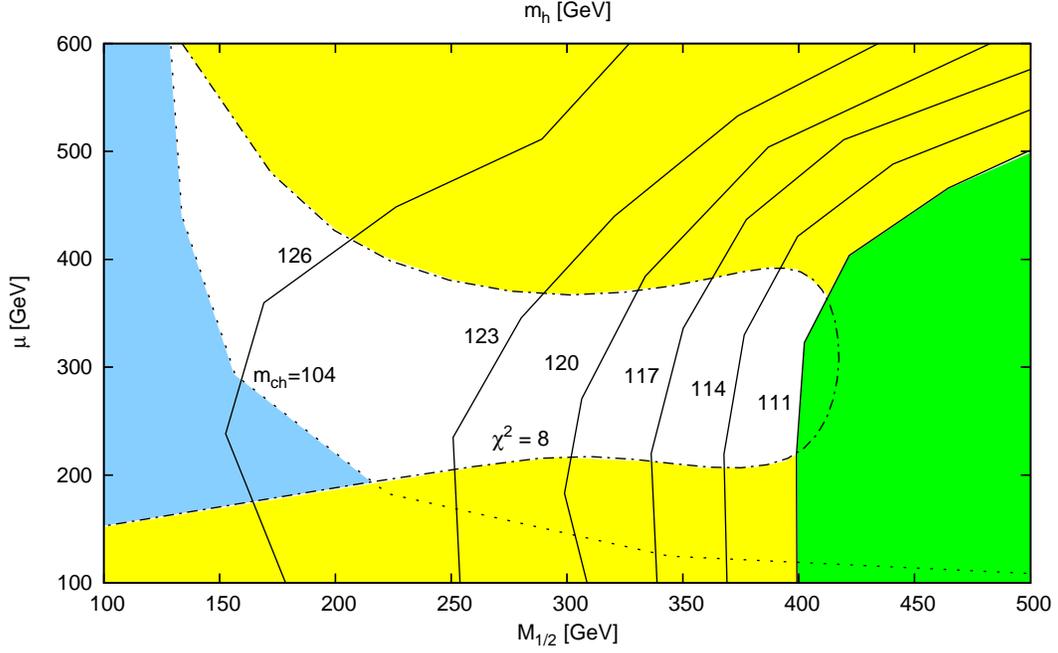,width=14cm} \\
\end{minipage}
\end{center}
\begin{center}
\caption{\label{fig:mh4tev} {\small Contours of constant light Higgs
    mass  $m_h$ for $m_{16} = 4$ TeV and $m_{A} = 700$ GeV. The shaded
    regions are the same as in Fig. 1.} }
\end{center}
\end{figure}

\begin{figure}[t!]
\begin{center}
\begin{minipage}{6in}
\hspace*{5mm}
\epsfig{figure=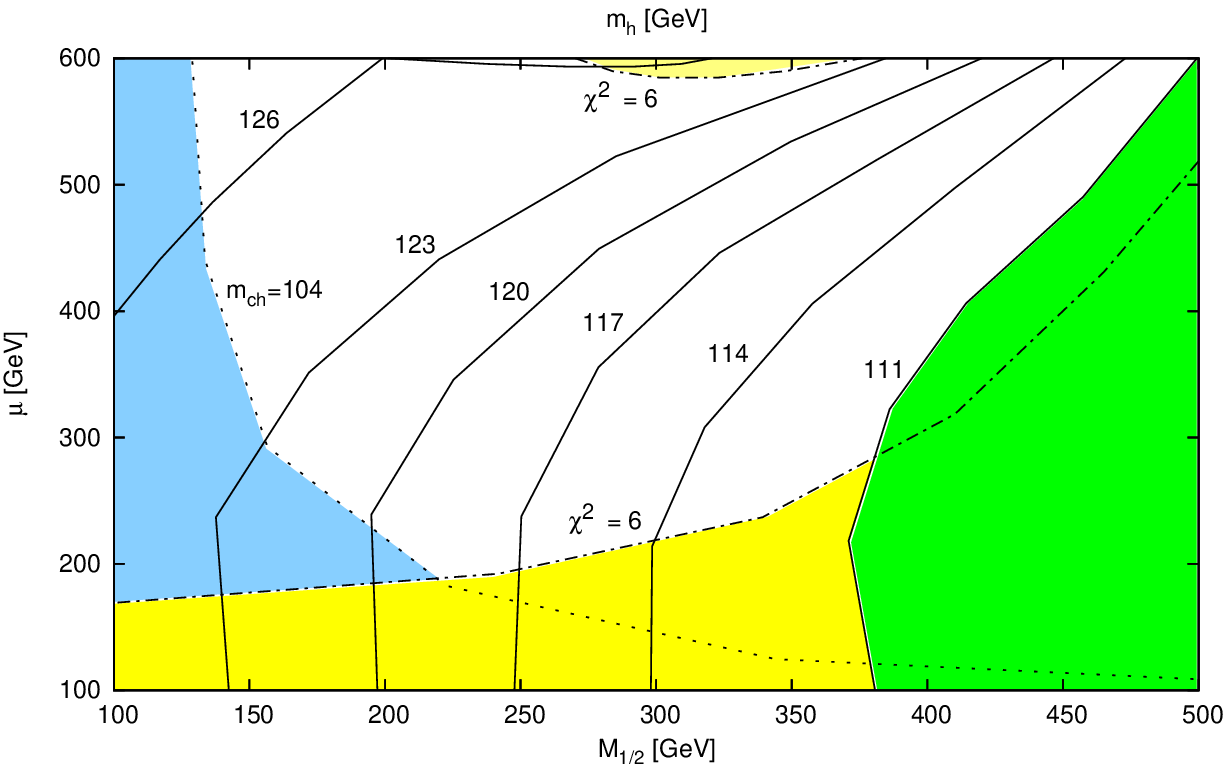,width=14cm} \\
\end{minipage}
\end{center}
\begin{center}
\caption{\label{fig:mh} {\small Contours of constant light Higgs
    mass  $m_h$ for $m_{16} = 5$ TeV and $m_{A} = 700$ GeV. The shaded
    regions are the same as in Fig. 2.} }
\end{center}
\end{figure}

In Figs. \ref{fig:mh4tev} and \ref{fig:mh} we present contours of
constant light Higgs mass for the case $m_{16} = 4$ and 5 TeV. There
is not much difference in the range of light Higgs mass in the two
cases.  We find an upper bound on the Higgs mass given by  $m_h \leq
127$ GeV.   In our analysis, we use the output of our RG running as
input to FeynHiggs \cite{Heinemeyer:1998yj} to obtain the Higgs pole
mass at two loops. Note, in the region of parameter space with
$|A_0|
> m_{16}$ the radiative corrections to the Higgs mass are
significant, i.e. the two loop correction, using FeynHiggs, is of
order 30 GeV.   One might then worry that the theoretical
uncertainty in the light Higgs mass is just as big. However,
Heinemeyer \cite{Heinemeyer:2004ms} (see sec 2.5), estimates the
uncertainties in the light Higgs mass from yet-to-be-calculated
two-loop corrections and higher to be at most 3 GeV.  We have thus
taken $\pm 3$ GeV as the estimated total theoretical uncertainty in
the light Higgs mass.\footnote{Our results for the light Higgs mass
are about 5 GeV higher than in our previous analyses
\cite{bdr,Dermisek:2003vn}. This is due to the fact that we now use
FeynHiggs to calculate the Higgs mass, where as in previous papers
we used an effective potential analysis. The disagreement of the
effective potential method with FeynHiggs (a perturbative analysis)
is well known, see for example \cite{Allanach:2004rh}. In addition
one sees that the light Higgs mass decreases as $M_{1/2}$ increases.
This fact is completely due to the global $\chi^2$ analysis and the
need to fit the bottom quark mass starting with 3rd family Yukawa
unification.  For a detailed discussion of this effect, see the
second reference in \cite{Dermisek:2003vn}.}


\subsection{Lepton flavor violation and electric dipole moments \label{sec:lfv}}
Let us now focus on our results for lepton flavor violating (LFV)
processes $l_j\to l_i\g$ and charged lepton electric dipole moments
(EDMs) in this theory.  We start with universal squark and slepton
masses and a universal A parameter at the GUT scale. This is thus a
GUT version of minimal flavor violating boundary conditions, giving
minimal flavor violation at low energies.  Thus the dominant
contribution to lepton flavor violation results from the RG running
of slepton masses and the effect of neutrino Yukawa couplings on
this running from the GUT scale to the heaviest right-handed
neutrino Majorana mass of order $10^{14}$ GeV. See for example the
seminal paper on this subject \cite{Borzumati:1986qx}. There is also
ample literature regarding LFV and EDMs, see for instance
\cite{Hisano:1995cp,Hall:1985dx} for LFV and
\cite{Ellis:2001xt,Dimopoulos:1994gj} for EDMs. Therefore we simply
quote the results of \cite{Hisano:1995cp} and \cite{Ellis:2001xt}
here and refer the reader to those references for more detail.

\begin{figure}[h!]
\begin{center}
\begin{tabular}{cc}
\epsfig{figure=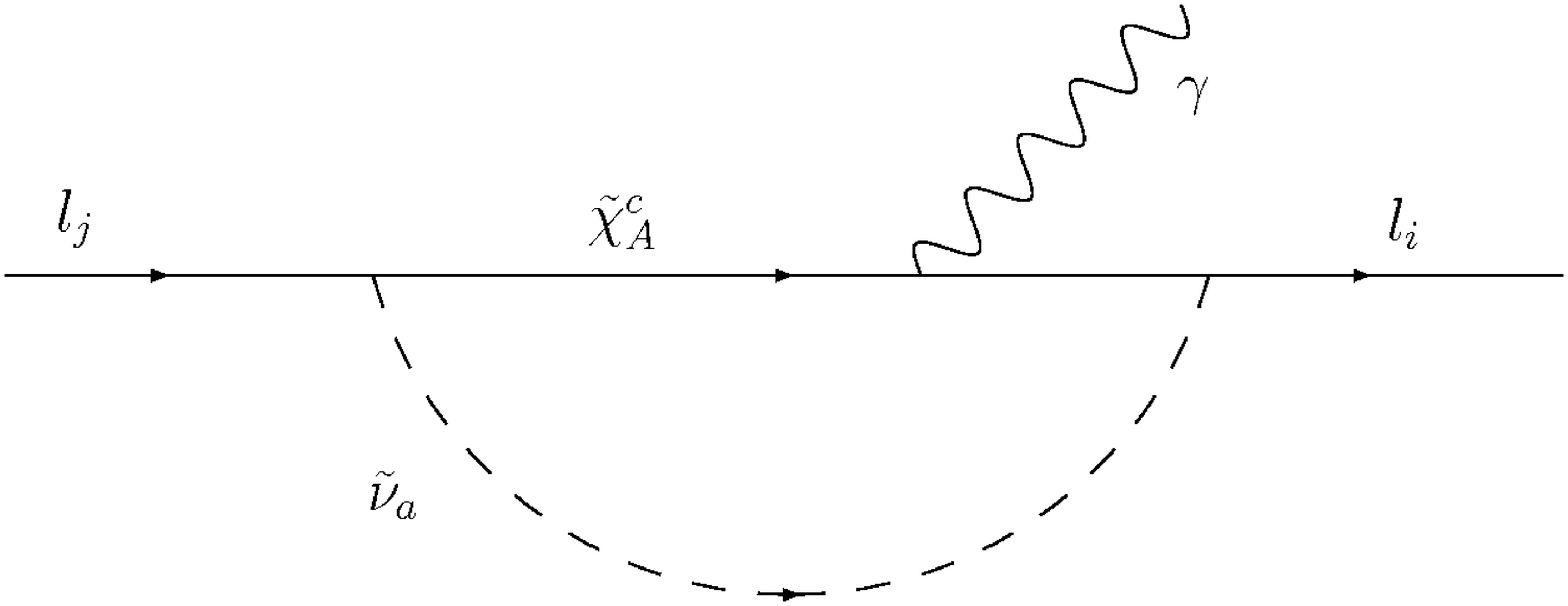,width=7cm}&
\epsfig{figure=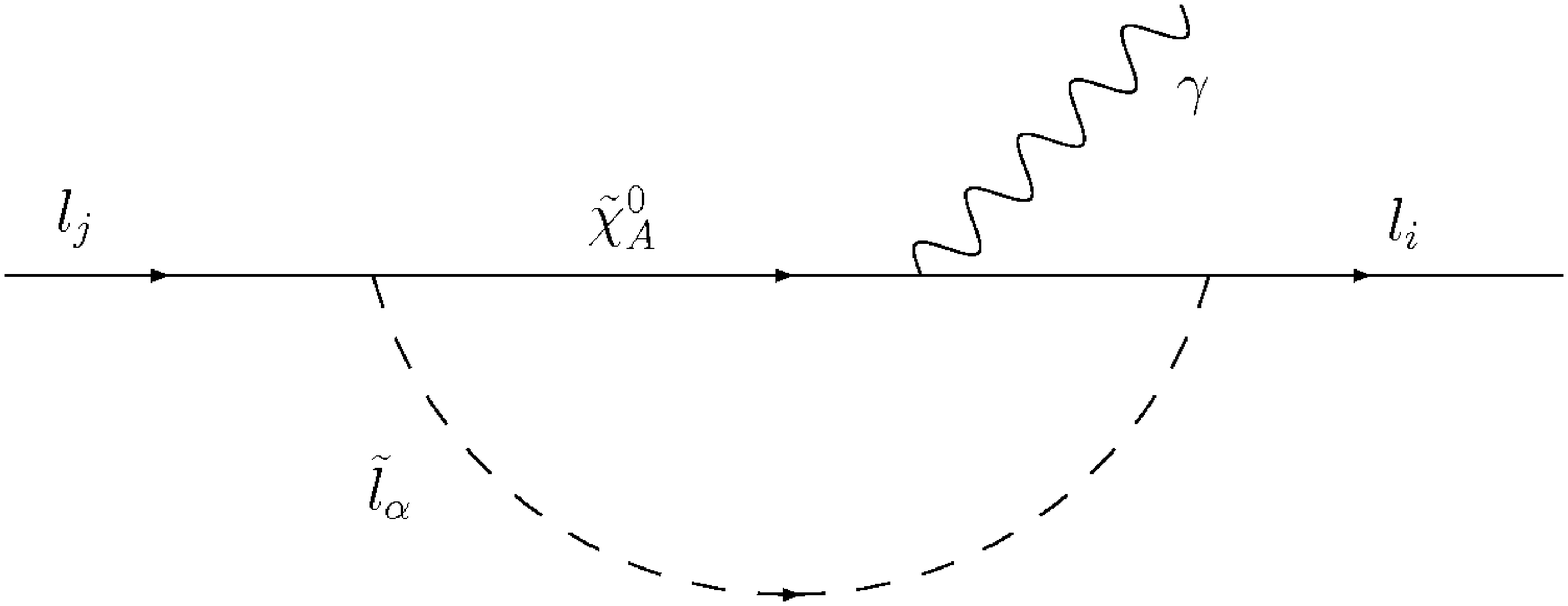,width=7cm} \\
(a)&(b)
\end{tabular}
\end{center}
\begin{center}
\caption{\label{fig:amps} {\small One-loop Feynman diagrams relevant
for the decay of $l_j\to l_i\g$. (a) involves charginos
$\tilde{\chi}^c_A$ and sneutrinos $\tilde{\nu}_{a}$, and (b)
involves neutralinos $\tilde{\chi}^0_A$ and sleptons $\tilde l_{\a}
\ $ in the loop with $A=1,2$ for charginos, $A=1,2,3,4$ for
neutralinos, $\ a = 1,2,3$ and $\a=1,\ldots,6$}. }
\end{center}
\end{figure}

Following the notation of \cite{Hisano:1995cp}, the effective
Lagrangian ${\cal L}$ relevant for the decay $l_j\to l_i\g$ is
\[ {\cal L}=-\frac e2 m_{l_j}\bar u_i\sigma_{\a \beta}
    (A_2^LP_L+A^R_2P_R)u_jF^{\a\beta}, \]
where $e$ is the electric charge, $m_{l_j}$ is the mass of the
decaying lepton, $P_{R/L}$ is the chirality projection operator,
$u_i$ and $u_j$ are Dirac spinors describing $l_i$ and $l_j$,
respectively. $A_2^{L/R}$ is obtained by calculating Feynman
diagrams depicted in Figure \ref{fig:amps} at one loop, and found in
\cite{Hisano:1995cp}.

The decay amplitude is given by
\begin{equation*}
    T=ie\ep^{\a *} m_{l_j}\bar u_i(p-q)\sigma_{\a\beta}q^{\beta}
    (A_2^LP_L+A^R_2P_R)u_j(p),
\end{equation*}
where $\ep^*$ is the photon polarization vector. Then, the decay
rate is
\[  \Gamma(l_j\to l_i\g)=\frac {e^2}{16\pi}m_{l_j}^5
    (|A_2^L|^2+|A_2^R|^2). \]

\begin{figure}[h!]
\begin{center}
\begin{tabular}{cc}
\begin{minipage}{4cm}
\epsfig{figure=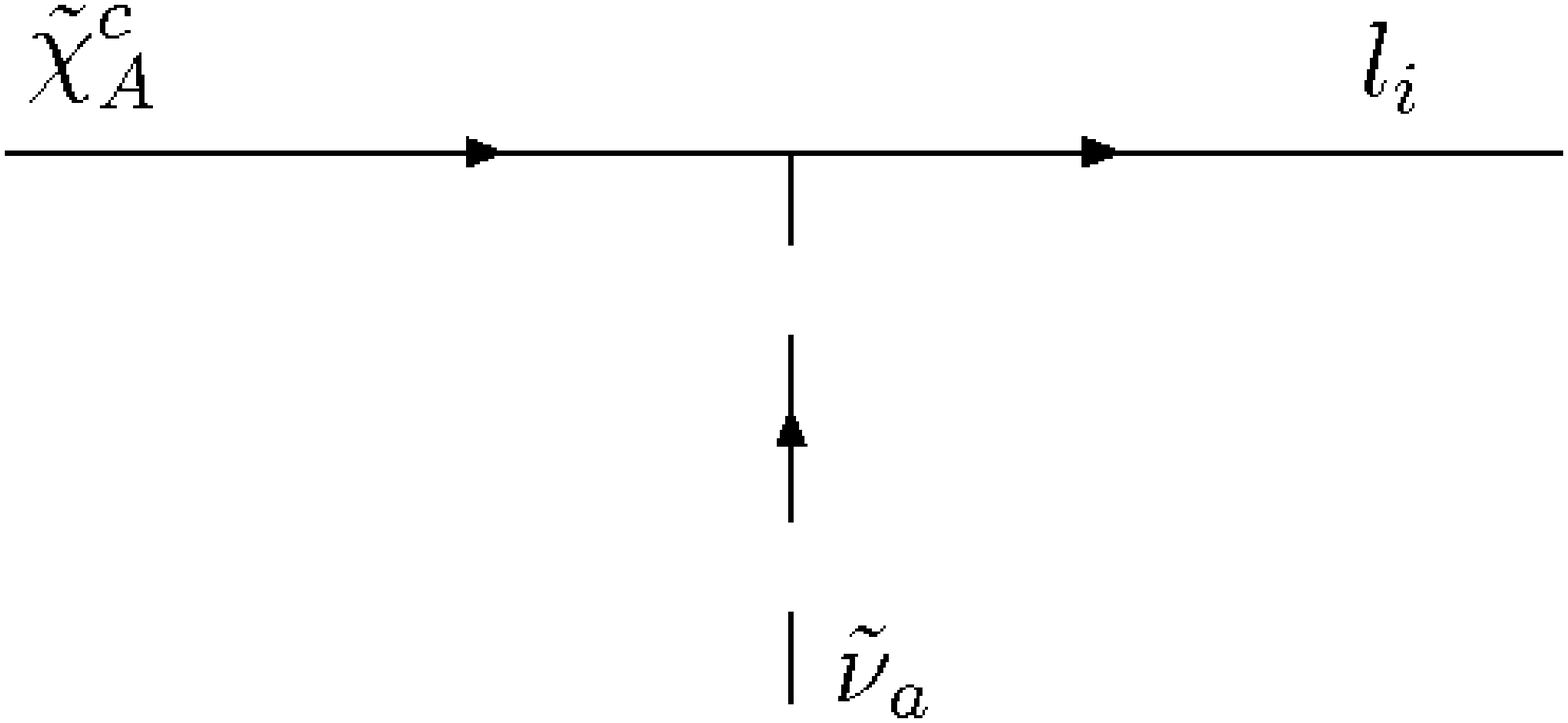,width=4cm}
\end{minipage}
{\scriptsize $\equiv i(C^{R(l)}_{iAa}P_R+C^{L(l)}_{iAa}P_L)$}&
\begin{minipage}{4cm}
\epsfig{figure=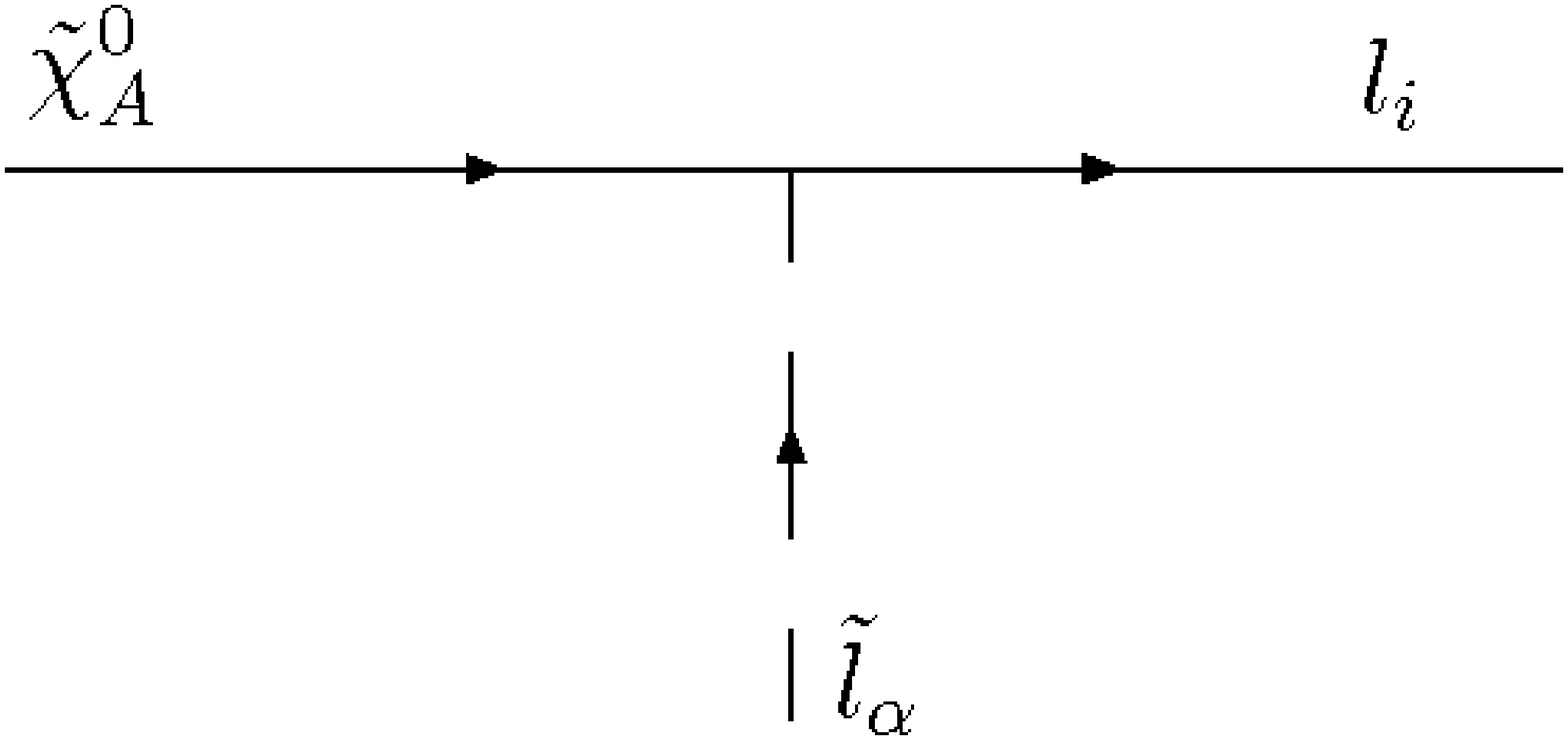,width=4cm}
\end{minipage}
{\scriptsize $\equiv i(N^{R(l)}_{iA\a}P_R+N^{L(l)}_{iA\a}P_L)$} \\
(a)&(b)
\end{tabular}
\end{center}
\begin{center}
\caption{\label{fig:vers} {\small Vertices (a) define
$C_{iAa}^{L(l)}$ and $C_{iAa}^{R(l)}$ and vertices (b)
$N_{iA\a}^{L(l)}$ and $N_{iA\a}^{R(l)}$ with $A=1,2$ for charginos,
$A=1,2,3,4$ for neutralinos, and $\a=1,\ldots,6$}. }
\end{center}
\end{figure}

On the other hand, the lepton electric dipole moment $d_{l_i}$ is
defined as the coefficient of the effective Lagrangian ${\cal L}$ of
the form
\[ {\cal L}=-\frac i2 d_{l_i} \bar u_i\sigma_{\a\beta}
    \g_5u_iF^{\a\beta}.\] Let us write
\[ d_{l_i}\equiv d^{ch}_{l_i}+d^{nt}_{l_i},\]
where $d^{ch}_{l_i}$ and $d^{nt}_{l_i}$ are contributions to the EDM
from loops in Figure \ref{fig:amps} (a) and (b) with replacing $l_j$
by $l_i$, respectively. Then we find
\begin{gather*}
    d_{l_i}^{ch}=-\frac{e}{16\pi^2}\sum_{A=1}^2\sum_{a=1}^3{\rm Im}
    (C_{iAa}^{L(l)}C_{iAa}^{R(l)*})\frac{m_{\tilde
    \chi_A^+}}{m_{\tilde{\nu}_a}^2}\frac{3-4x_{Aa}
    +x_{Aa}^2+2\ln x_{Aa}}{2(1-x_{Aa})^3},\\
    d_{l_i}^{nt}=-\frac{e}{16\pi^2}\sum_{A=1}^4\sum_{\a=1}^6{\rm Im}
    (N_{iA\a}^{L(l)}N_{iA\a}^{R(l)*})\frac{m_{\tilde
    \chi_A^0}}{m_{\tilde{l}_{\a}}^2}\frac{1-x_{A\a}^2+2x_{A\a}
    \ln x_{A\a}}{2(1-x_{A\a})^3}. \end{gather*}
Here $C_{iAa}^{L(l)}$, $C_{iAa}^{R(l)}$, $N_{iA\a}^{L(l)}$ and
$N_{iA\a}^{R(l)}$ are read from vertices shown in Figure
\ref{fig:vers} and the expression for them is given in
\cite{Hisano:1995cp}. $m_{\tilde \chi_A^+}$ is chargino mass,
$m_{\tilde{\nu}_a}$ is sneutrino mass, $m_{\tilde \chi_A^0}$ is
neutralino mass, $m_{\tilde{l}_{\a}}$ is selectron mass,
$x_{Aa}\equiv m_{\tilde{\chi}_A^+}^2/m_{\tilde {\nu}_{a}}^2$ and
$x_{A\a}\equiv m_{\tilde{\chi}^0_A}^2/m_{\tilde {l}_{\a}}^2$.

Table \ref{table:expts} shows the current limits on various lepton
flavor violating processes and EDMs with an estimated sensitivity
for future experiments. See for example \cite{Aoki:2005gw} for a
summary of the current and future experimental status on searches
for LFV and the muon EDM.

\begin{table}[ht]
\begin{center}
\begin{tabular}{|c|c|c|}
\hline
            &Current limit &Expected accuracy \\ \hline\hline
\hline
$Br(\mu\to e\g)$&
    $<1.2\times 10^{-11}$ \cite{Brooks:1999pu}&
    $5\times 10^{-14}$ \cite{Grassi:2005ac} \\
$Br(\tau\to e\g)$&
    $<1.1\times 10^{-7}$ \cite{Aubert:2005wa}&
    $-$ \\
$Br(\tau\to \mu\g)$&
    $<6.8\times 10^{-8}$ \cite{Aubert:2005ye}&
    $-$ \\
$Br(\mu\to 3e)$&
    $<1.0\times 10^{-12}$ \cite{Bellgardt:1987du}&
    $-$ \\
$Br(\tau\to 3l)$&
    $<1-3\times 10^{-7}$ \cite{Aubert:2003pc}&
    $-$ \\
$Br(\mu Ti\to e Ti)$&
    $<1.7\times 10^{-12}$ \cite{Kaulard:1998rb}&
    $10^{-18}$\cite{Kuno:2005ac} \\ \hline
$d_{e}$ [$e\cdot cm$] &
    $<1.6\times 10^{-27}$ \cite{Regan:2002ta}&
    $10^{-31}$ \cite{Kawall:2004nv}\\
$d_{\mu}$ [$e\cdot cm$] &
    $<10^{-18}$ \cite{Bailey:1978mn}&
    $10^{-24}-10^{-25}$ \cite{Farley:2003wt} \\
$d_{\tau}$ [$e\cdot cm$] &
    $<10^{-16}$ \cite{Inami:2002ah}&
    $-$\\
\hline
\end{tabular}
\end{center}
\caption{Current upper limit on various LFV processes and lepton
EDMs. The limits on the LFV processes are all at 90 \% CL. Estimated
accuracy for the future experiment is given for $\mu\to e\g$, $\mu
e$ conversion, electron EDM and muon EDM.} \label{table:expts}
\end{table}

We note that the MEG collaboration \cite{Grassi:2005ac} (searching
for $\mu\to e\g$ with a sensitivity of order $\geq 5\times
10^{-14}$) should start taking data by September 2006 and may have
significant results by 2008. It will be an excellent test for any
new physics beyond the standard model; and in particular, the DR
model \cite{Dermisek:2005ij}. Let us also note here that we can
calculate $Br(l_i\to 3l_j)$ to a good approximation by using
\[ \frac{Br(l_i\to 3l_j)}{Br(l_i\to l_j\g)}\simeq\frac {\a}{3\pi}
    \left(\log\frac{m_{l_i^2}}{m_{l_j}^2}-\frac{11}4\right),\]
which has been verified in Ref.~\cite{Arganda:2005ji} for all values
of $\tan\beta$. In particular
\[  \frac{Br(\tau\to 3\mu)}{Br(\tau\to \mu\g)}\simeq\frac 1{440},\quad
    \frac{Br(\tau\to 3e)}{Br(\tau\to e\g)}\simeq\frac 1{94},\quad
    \frac{Br(\mu\to 3e)}{Br(\mu\to e\g)}\simeq\frac 1{162}. \]
This means that if we satisfy the constraint from $Br(l_i\to
l_j\g)$, we automatically also satisfy the constraint from
$Br(l_i\to 3l_j)$.

\begin{figure}[ht]
\begin{center}
\begin{tabular}{c}
\epsfig{figure=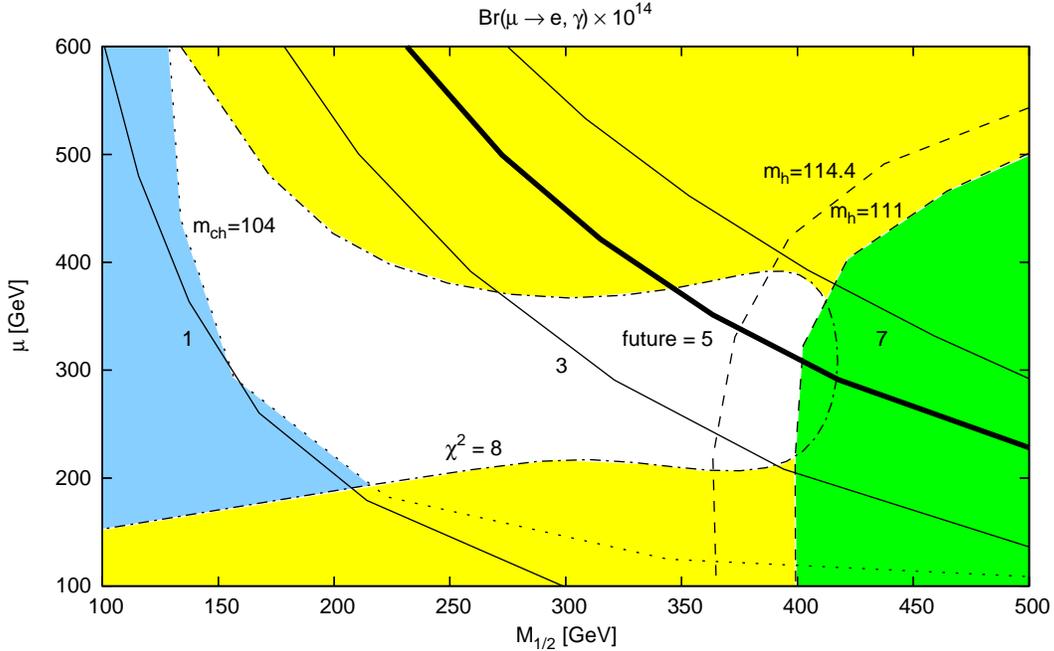,width=14cm}
\end{tabular}
\caption{\label{fig:muegamma4tev} {\small Contours of constant
branching
    ratio $Br(\mu \rightarrow e \gamma) \times 10^{14}$ for $m_{16} = 4$ TeV and $m_{A} = 700$ GeV. The shaded
    regions are the same as in Fig. 1.  Note the narrow region in the upper right hand corner with $\chi^2 \leq 8$ which
    is within the sensitivity of the MEG experiment. } }
\end{center}
\end{figure}

\begin{figure}[ht]
\begin{center}
\begin{tabular}{c}
\epsfig{figure=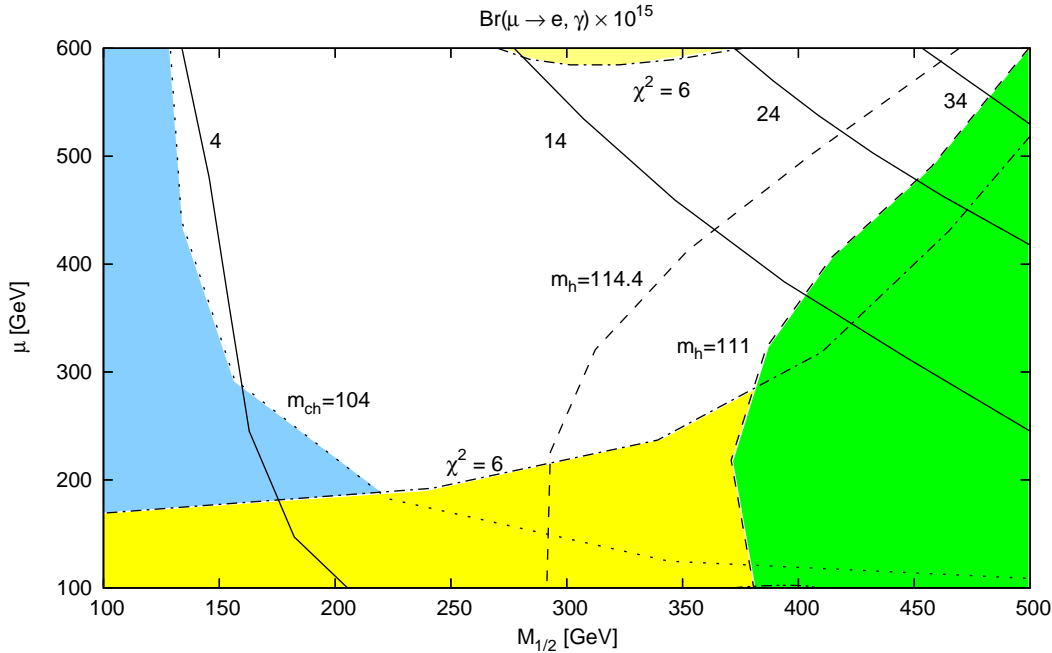,width=14cm}
\end{tabular}
\caption{\label{fig:muegamma} {\small Contours of constant branching
    ratio $Br(\mu \rightarrow e \gamma) \times 10^{15}$ for $m_{16} = 5$ TeV and $m_{A} = 700$ GeV. The shaded
    regions are the same as in Fig. 2.  Note all points are below the sensitivity of the MEG
    experiment. } }
\end{center}
\end{figure}

\begin{figure}[t!]
\begin{center}
\begin{minipage}{6in}
\hspace*{12mm} \epsfig{figure=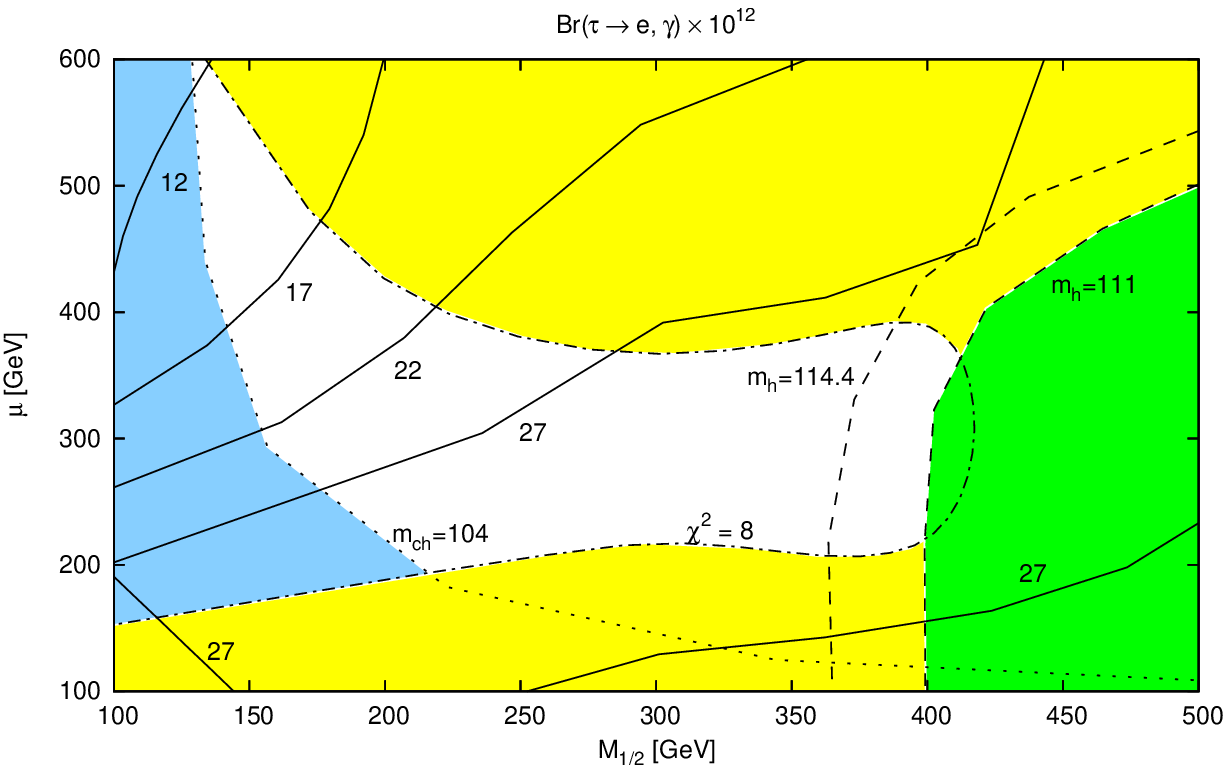,width=12cm}
\end{minipage}
\begin{minipage}{6in}
\hspace*{12mm} \epsfig{figure=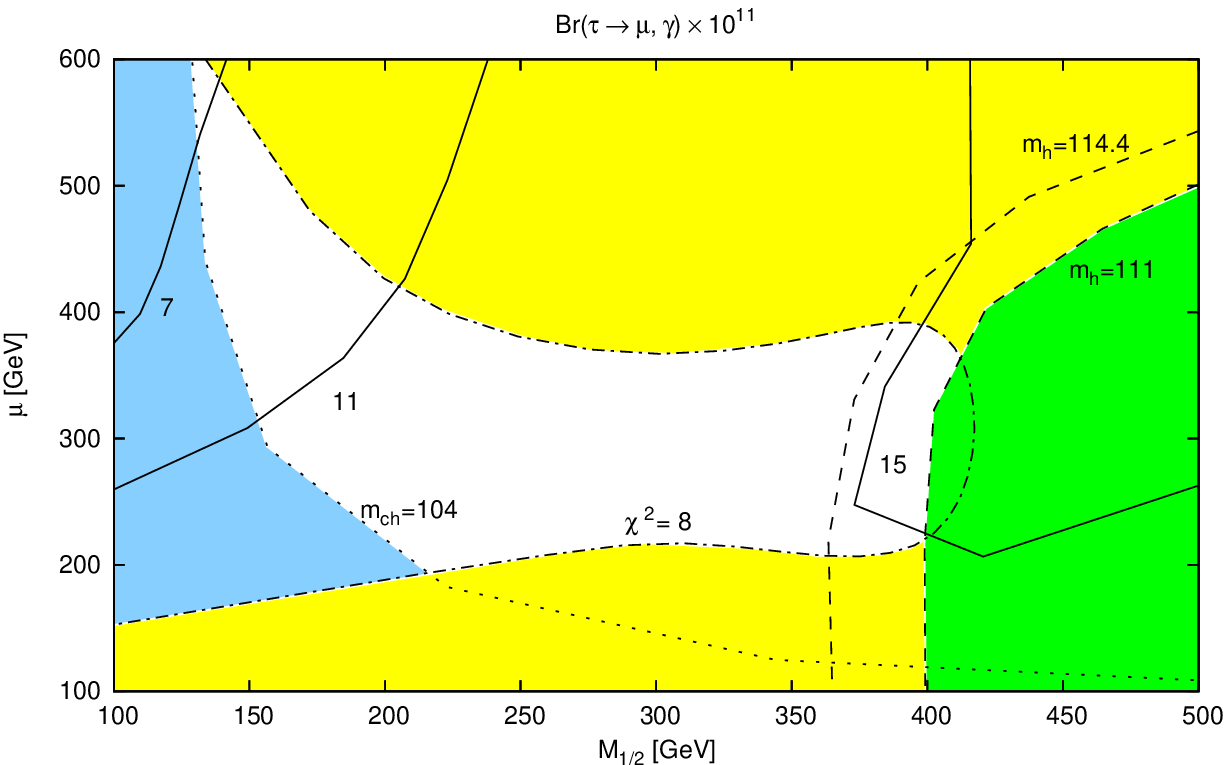,width=12cm}
\end{minipage}
\end{center}
\begin{center}
\caption{\label{fig:tau2lgamma} {\small Contours of constant
    branching ratio $Br(\tau \rightarrow e \gamma) \times 10^{12}$ (upper) and  $Br(\tau \rightarrow \mu \gamma)
    \times 10^{11}$ (lower)
    for $m_{16} = 4$ TeV and $m_{A} = 700$ GeV. The shaded regions are the same as in Fig. 1. } }
\end{center}
\end{figure}

In Fig. \ref{fig:muegamma4tev} we plot contours of constant
branching ratio $Br(\mu \rightarrow e \gamma)$ for $m_{16} = 4$ TeV.
The prediction is significantly below the present experimental
bounds.  Moreover, comparing this result with the future sensitivity
of the MEG experiment \cite{Grassi:2005ac} ($Br(\mu \rightarrow e
\gamma) > 5 \times 10^{-14}$) we find that our prediction is below
the MEG sensitivity in most of the parameter space.  Note, however,
the narrow region in the upper right hand corner with $\chi^2 \leq
8$ which is within the sensitivity of the MEG experiment.  In Fig.
\ref{fig:muegamma}  we present results for $m_{16} = 5$ TeV.  Of
course,  larger scalar masses suppress the branching ratio, so that
now the entire allowed region is below the projected MEG
sensitivity.    The results for the decays $\tau \rightarrow e
\gamma$ and $\tau \rightarrow \mu \gamma$ are given in Fig.
\ref{fig:tau2lgamma}.   Unfortunately the results are significantly
below the present bounds and we are not aware of any experiments to
significantly improve these bounds.

\begin{figure}[ht]
\begin{center}
\begin{tabular}{c}
\epsfig{figure=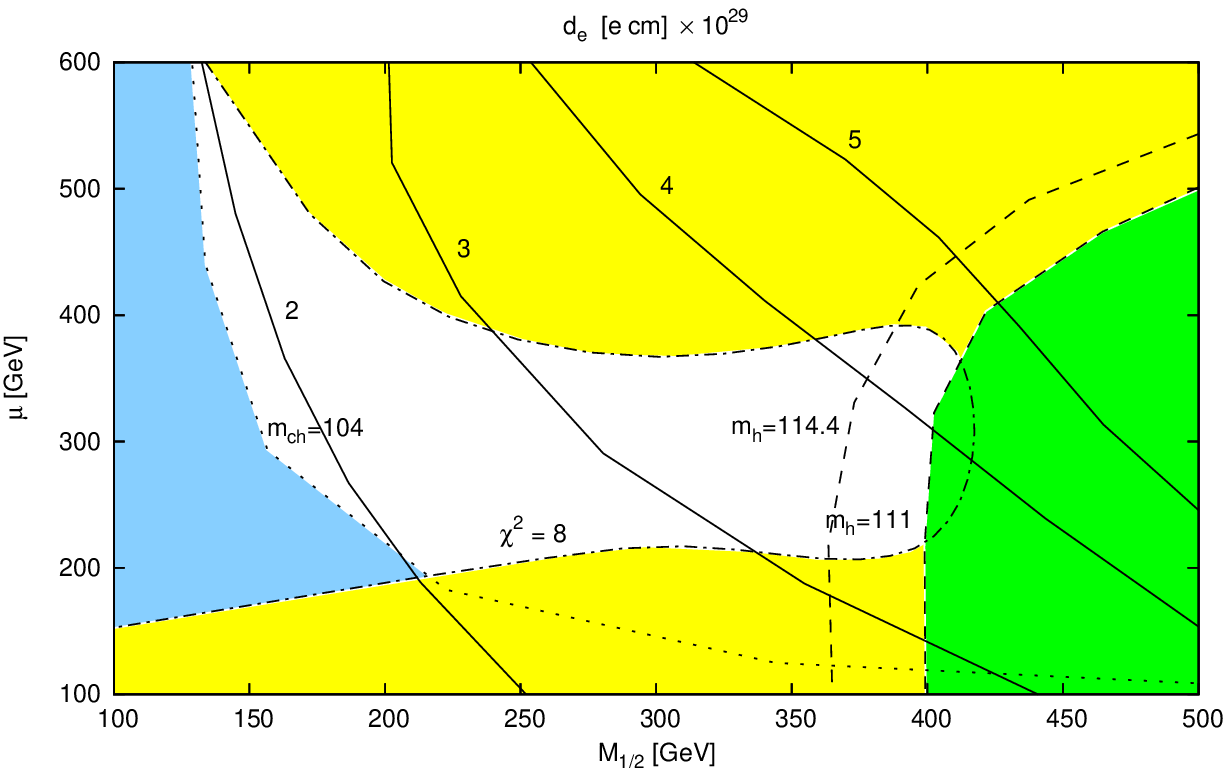,width=14cm}
\end{tabular}
\caption{\label{fig:edmE4tev} {\small Contours of constant electric
    dipole moment of the electron, $d_e [e \cdot cm]\times 10^{29}$, for $m_{16} = 4$ TeV and $m_{A}= 700$ GeV. The shaded
    regions are the same as in Fig. 1.  Note the entire region is within the sensitivity of future
    experiments.} }
\end{center}
\end{figure}

\begin{figure}[ht]
\begin{center}
\begin{tabular}{c}
\epsfig{figure=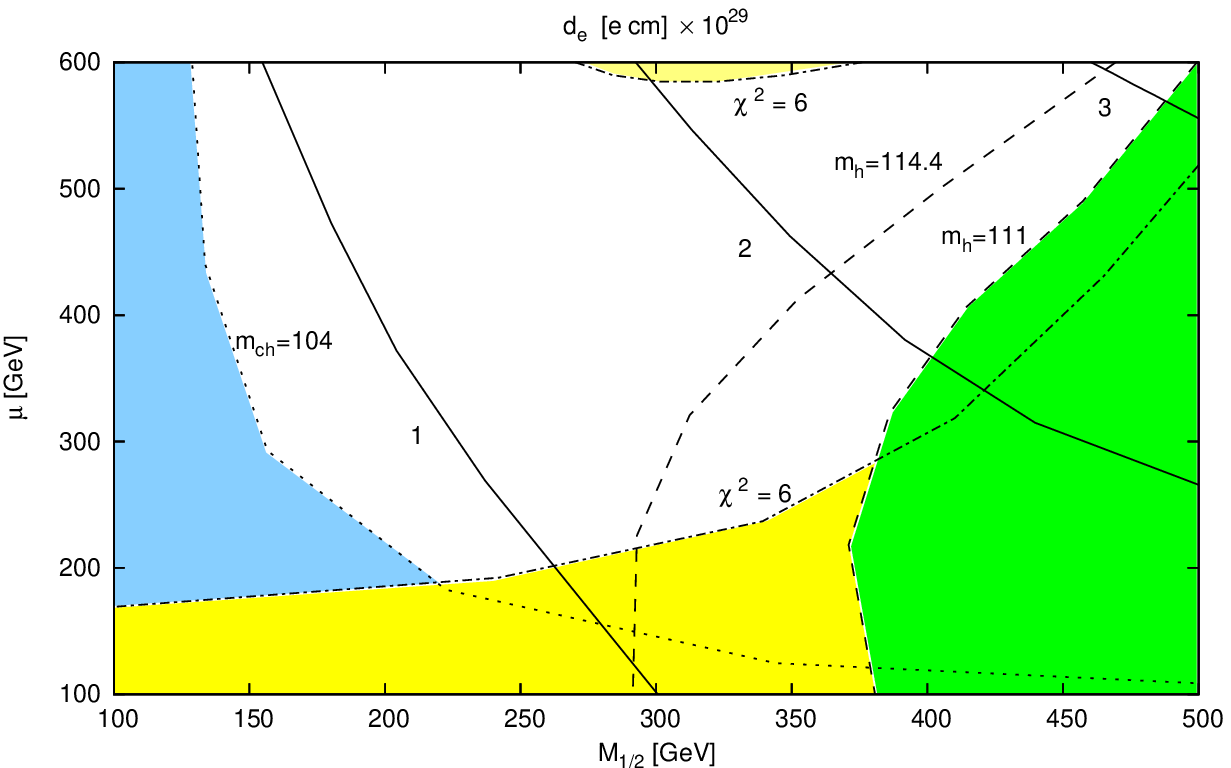,width=14cm}
\end{tabular}
\caption{\label{fig:edmE} {\small Contours of constant electric
    dipole moment of the electron, $d_e [e \cdot cm] \times 10^{29}$, for $m_{16} = 5$ TeV and $m_{A}= 700$ GeV. The shaded
    regions are the same as in Fig. 2.  Note the entire region is within the sensitivity of future
    experiments.} }
\end{center}
\end{figure}

\begin{figure}[t!]
\begin{center}
\begin{minipage}{6in}
\hspace*{12mm} \epsfig{figure=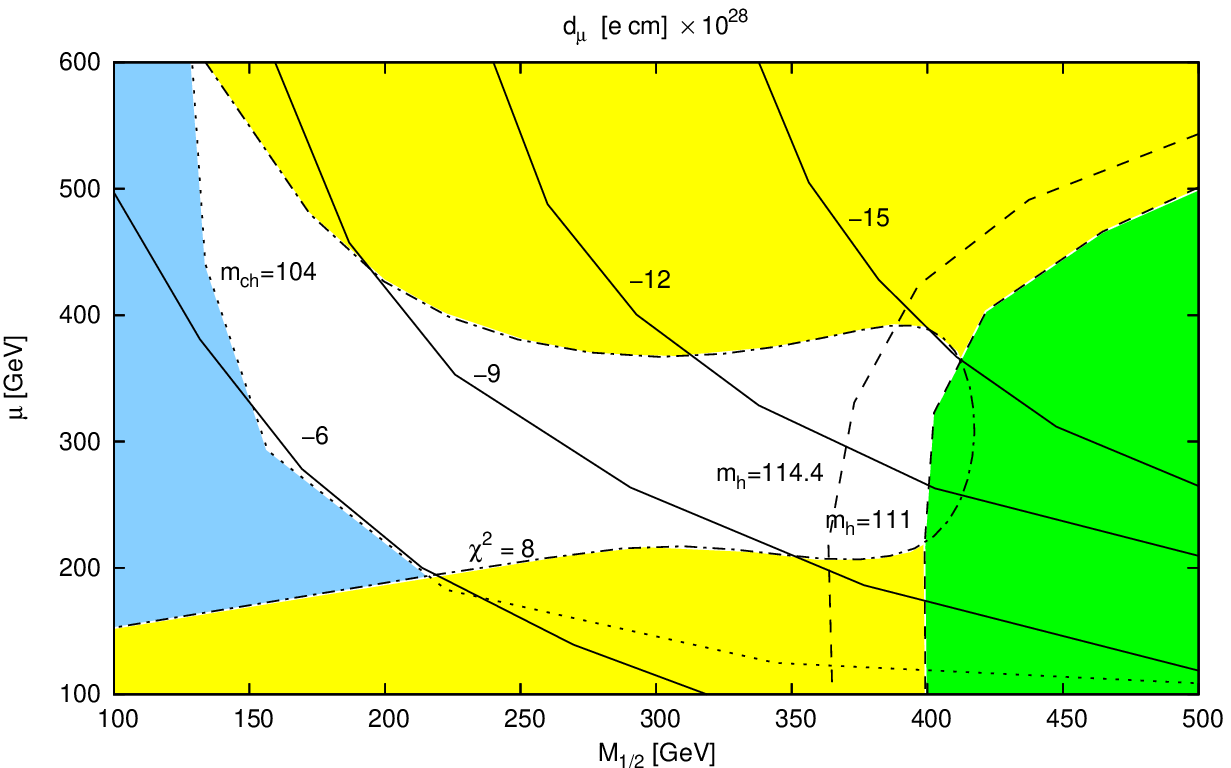,width=12cm}
\end{minipage}
\begin{minipage}{6in}
\hspace*{12mm} \epsfig{figure=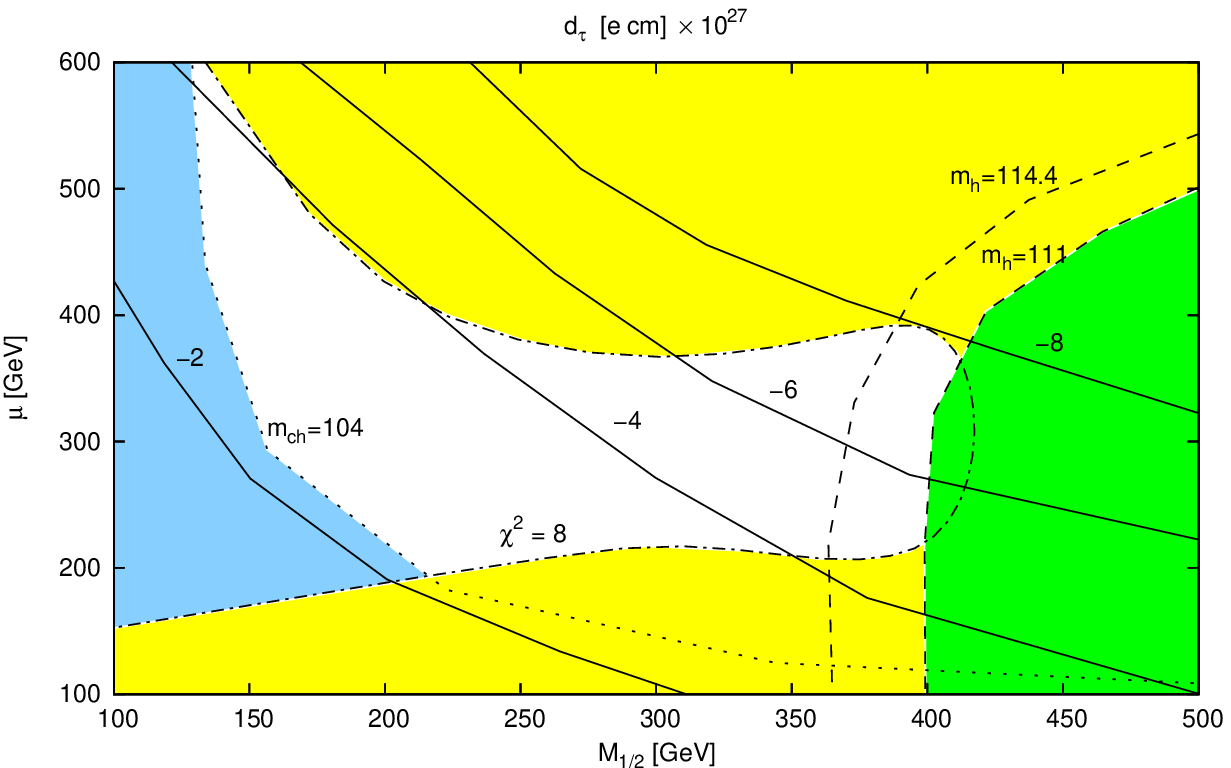,width=12cm}
\end{minipage}
\end{center}
\begin{center}
\caption{\label{fig:edmMT} {\small Contours of constant electric
    dipole moment of the muon, $d_\mu [e \cdot cm] \times 10^{28}$, (upper) and the tau, $d_\tau [e \cdot cm] \times 10^{27}$, (lower) for
    $m_{16} = 4$ TeV and $m_{A}= 700$ GeV. The shaded regions are the same as in Fig. 1.
    Note, the entire region is below present bounds and for $d_\mu$
the result is below the projected
    sensitivity of future experiments. } }
\end{center}
\end{figure}

We have also evaluated the predictions for the electric dipole
moment of the electron, muon and tau.    In Figs. \ref{fig:edmE4tev}
and \ref{fig:edmE} we present the results for the electric dipole
moment of the electron.   Note, in both cases, the entire region is
below the present bounds, and also within the projected sensitivity
of future experiments \cite{Kawall:2004nv}.  In Fig. \ref{fig:edmMT}
we present the results for the electric dipole moments of the muon
and tau.   In all cases the results are below the present bounds and
for $d_\mu$ the result is below the projected sensitivity of future
experiments \cite{Farley:2003wt}.

\begin{figure}[ht]
\begin{center}
\begin{tabular}{c}
\epsfig{figure=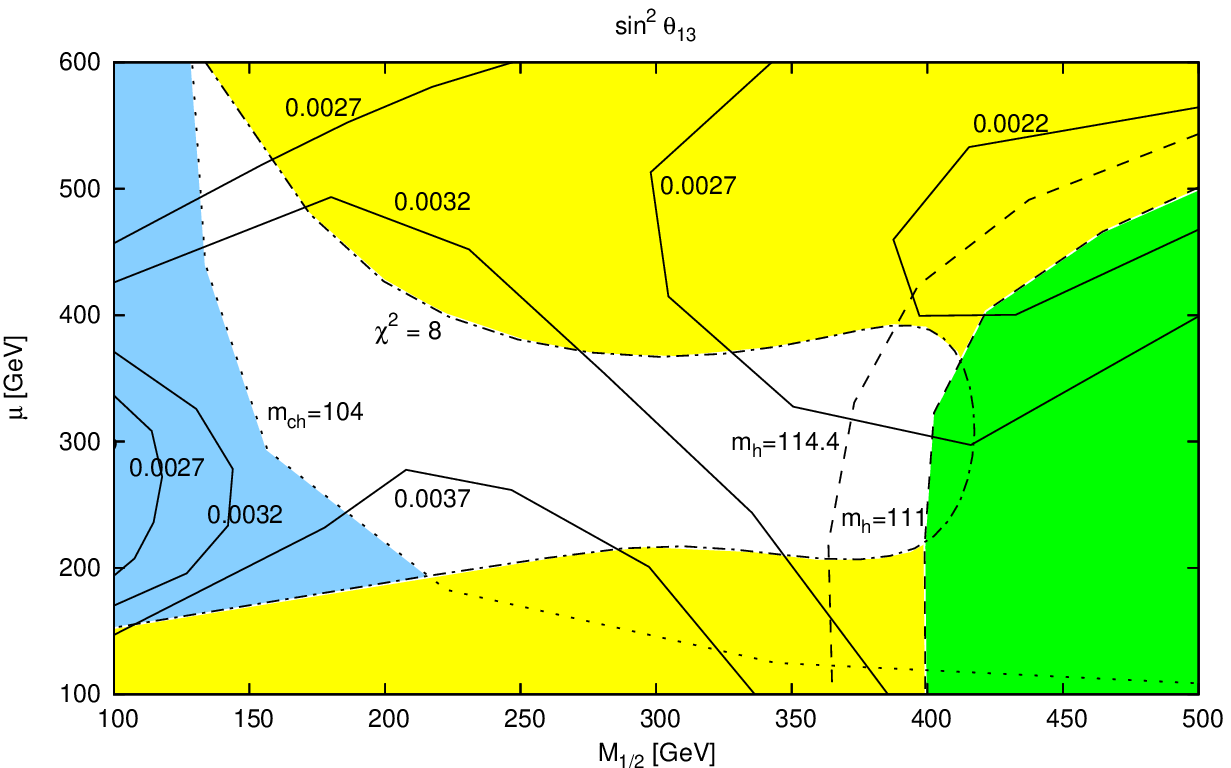,width=14cm}
\end{tabular}
\caption{\label{fig:s13.4tev} {\small Contours of constant neutrino
      mixing angle, $\sin^2\theta_{13}$, for $m_{16} = 4$ TeV and $m_{A}= 700$ GeV. The
      shaded regions are the same as in Fig. 1.   } }
\end{center}
\end{figure}

\begin{figure}[ht]
\begin{center}
\begin{tabular}{c}
\epsfig{figure=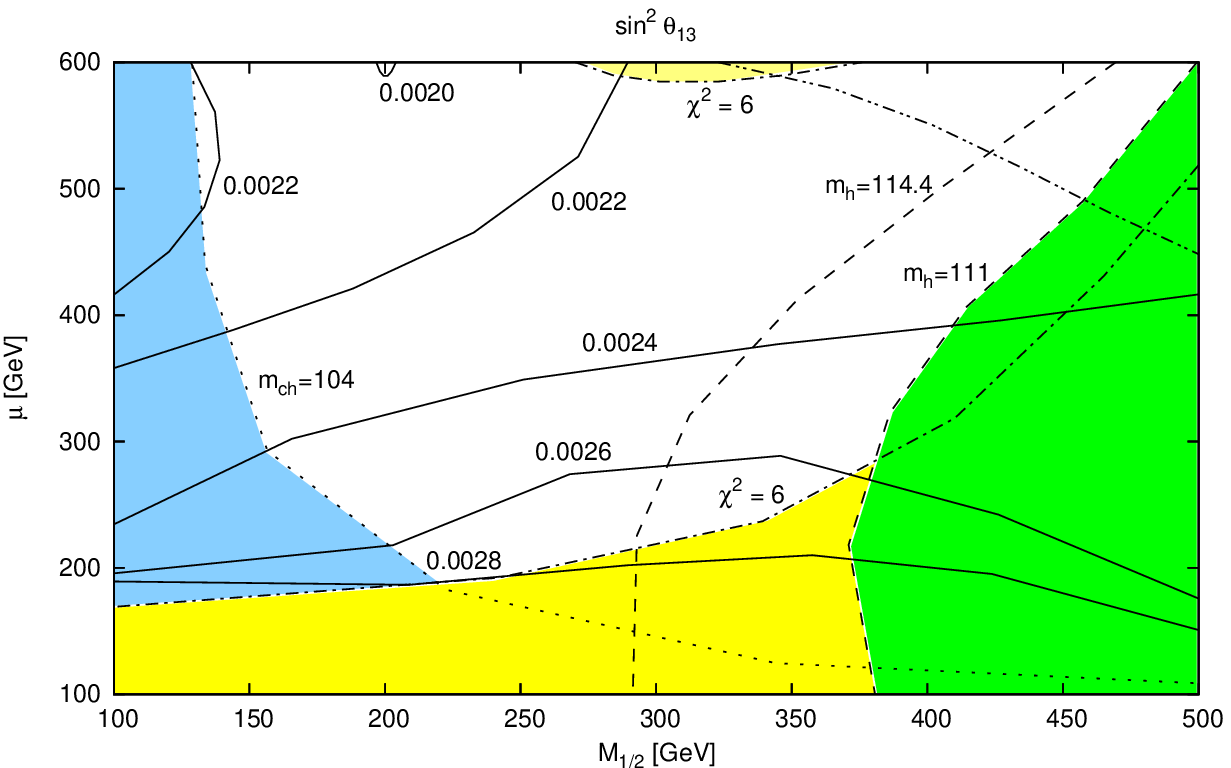,width=14cm}
\end{tabular}
\caption{\label{fig:s13} {\small Contours of constant neutrino
      mixing angle, $\sin^2\theta_{13}$, for $m_{16} = 5$ TeV and $m_{A}= 700$ GeV. The
      shaded regions are the same as in Fig. 2.   } }
\end{center}
\end{figure}

In Figs. \ref{fig:s13.4tev} and \ref{fig:s13} we evaluate the
neutrino mixing angle $\sin^2\theta_{13}$.  Recall that measuring
this mixing angle is the goal of several future reactor and long
baseline neutrino experiments. Moreover, a sufficiently large value
for $\sin^2\theta_{13}$ is needed in order to have the possibility
of observing CP violation in neutrino oscillations.   The value of
$\sin^2 \theta_{13}$ is somewhat sensitive to the value of $m_{16}$;
with a central value changing from $\sin^2 \theta_{13} \sim 0.0030
\pm 0.0007$ for $m_{16} = 4$ TeV to $\sin^2 \theta_{13} \sim 0.0024
\pm 0.0004$ for $m_{16} = 5$ TeV (where the uncertainty corresponds
to varying over the range for $\mu, \ M_{1/2}$ with $\chi^2 \leq 8$
or $\leq 6$ in the two cases). Note, while $\sin^2 \theta_{13}$ is
relatively insensitive to varying $\mu, \ M_{1/2}$; the CP violating
parameter $\sin\delta$, on the other hand, is quite sensitive. We
find that $\sin\delta$ can vary between 0.1 and 1.0 for different
values of $\mu, \ M_{1/2}$, and as a result the CP violating
Jarlskog parameter J ranges between 0.0013 and 0.013.   CP violation
in the latter case may be observable at long baseline experiments.
For example, the JPARC-SK experiment has a potential sensitivity to
$\sin^2 2\theta_{13} < 1.5\times 10^{-3}$ and $\delta \sim \pm
20^\circ$ and a comparable sensitivity is expected from the
``Off-axis NUMI" proposal~\cite{McKeown:2004yq} .

In Tables \ref{t:fit4tev} and \ref{t:fit} we present the $\chi^2$
fit for a particular point in SUSY parameter space for $m_{16} = 4$
and 5 TeV, respectively. The points give a value of $\chi^2 = 7.65$
and 4.99.  The former is acceptable while the latter is quite good.
In the table caption we present the input data at the GUT scale. We
also show the heavy Majorana neutrino masses (roughly $10^{10},
10^{12}, 10^{14}$ GeV) responsible for the See-Saw mechanism and the
light neutrino masses.

Note, that the pull from $m_b$ and $M_b - M_c$ is significantly
lower for $m_{16} = 5$ TeV than for $m_{16} = 4$ TeV.   This
accentuates the ``tug of war" between the gluino and chargino loop
contributions to the bottom quark mass at large $\tan\beta$.   The
light quark mass ratio $m_d/m_s$ is difficult to fit and contributes
significantly to the pull in both cases.   This mass ratio is
particularly sensitive to the Georgi-Jarlskog ansatz relating first
and second generation quark and lepton masses.   The conflict here
is with the very low value of the strange quark mass, of order 105
MeV, preferred by lattice gauge theory calculations.   Finally,  we
note that in a previous analysis \cite{Dermisek:2005ij} $\sin2\beta$
contributed a value of 1.5 to the pull.   However the recent Belle
data gives a significantly smaller central value for $\sin2\beta$
and now the fit is significantly improved.

We present the additional predictions for squark, slepton and Higgs
masses, at these two points in SUSY parameter space, in
Table~\ref{t:susy_and_higgs}, and for neutrino masses and mixing
parameters, in Table~\ref{t:neutrino_predictions}.    We have given
the value for the effective mass parameter observable in
neutrinoless double beta decay \begin{eqnarray} \left< m_{\beta
\beta} \right> = & \left| \ \sum_i U_{ei}^2 \ m_{\nu_i} \ \right| =
& \left| \ \sum_i \left| U_{ei} \right|^2 \ m_{\nu_i} \ e^{i
\alpha'_i} \ \right| \end{eqnarray} (where $\alpha'_i = \alpha_i + 2
\delta, \;\; i=1,2$~\cite{Bilenky:2001rz}).  It is predicted to be
of order $2 \times 10^{-4}$ eV which is too low to see in
near-future experiments~\cite{Pascoli:2003ke,McKeown:2004yq}. We
also give the effective electron-neutrino mass observable, relevant
for the analysis of the low energy beta decay of tritium. This mass
parameter is unaffected by Majorana phases and is predicted to be an
order of magnitude larger. The observable,
\begin{equation}
m_{\nu_e}^{eff} = \left( \sum_i \left| U_{ei} \right|^2 m_{\nu_i}^2
\right)^{1/2}
\end{equation}
is predicted to be of order $6 \times 10^{-3}$ eV. The current
experimental limit is $m_{\nu_e}^{eff} \leq 2.5$ eV with the
possibility of future experiments, such as KATRIN, reaching bounds
on the order of 0.35 eV \cite{McKeown:2004yq}.  Unfortunately, both
mass parameters may be unobservable by presently proposed
experiments.   Finally, in Table~\ref{t:lfv_and_edm}, we present the
predictions for lepton flavor violation and the electric dipole
moments at the same points in SUSY parameter space.
\begin{table}
\caption[8]{ {\bf The fit for for fermion masses and mixing angles
at one particular point in SUSY parameter space defined by $m_{16} =
4$ TeV, $\mu = 300$ GeV and $M_{1/2} = 200$ GeV. } \\
   \mbox{Initial parameters: }\ \ \
\ \ (1/$\alpha_G, \, M_G, \, \epsilon_3$) = ($24.84,
\, 3.30 \times 10^{16}$ GeV, $\, -3.57$ \%), \makebox[1.8em]{ }\\
($\lambda, \, \lambda \epsilon, \, \sigma, \, \lambda \tilde \epsilon, \, \rho, \,
\lambda \epsilon', \, \lambda \epsilon \xi$) = ($ 0.62, \, 0.030, \, 0.87, \, 0.0063, \,
-0.059, \, -0.0021, \,
0.0040  $),\\
($\Phi_\sigma, \, \Phi_{\tilde \epsilon}, \, \Phi_\rho, \,
\Phi_\xi$) =  ($0.637, \, 0.453, \, 0.709, \, 3.609$) rad,
\makebox[6.6em]{ }\\
($m_{16}, \, M_{1/2}, \, A_0, \, \mu(M_Z)$) = ($4000,\, 200, \,
-7809.1, \,
300$) GeV,\\
($(m_{H_d}/m_{16})^2, \, (m_{H_u}/m_{16})^2, \, \tan\beta$) = ($1.91,  \, 1.61, \, 50.34$) \\
($M_{R_3}, \, M_{R_2}, \, M_{R_1}$) = ($4.6 \times 10^{13}$ GeV, $\,
- 8.1 \times 10^{11} $ GeV, $\, 1.1 \times
10^{10} $ GeV) \\ 
} \label{t:fit4tev}
$$
\begin{array}{|l|c|l|r|}
\hline
{\rm Observable}  &{\rm Data} \ (\sigma) & {\rm Theory} &  {\rm Pull} \\
{\rm (masses \ in \ GeV)}   &  &   &  \\
\hline
\;\;\;G_{\mu} \times 10^5   &  1.16637 \ (0.1 \%) & 1.16638  & < 0.01    \\
\;\;\;\alpha_{EM}^{-1} &  137.036 \ (0.1 \%) & 137.035  &  < 0.01        \\
\;\;\;\alpha_s(M_Z)    &  0.1187 \ (0.002) &   0.1174 &  0.37       \\
\hline
\;\;\;M_t              &  172.7 \ (2.9)   &   173.11  &       0.02 \\
\;\;\;m_b(M_b)          &    4.25 \ (0.25) &  4.49    &   0.94               \\
\;\;\;M_b - M_c        &    3.4 \ (0.2) &   3.61  &  {\bf 1.16}              \\
\;\;\;m_c(m_c)  &   1.2 \ (0.2)   &   1.16    &      0.03      \\
\;\;\;m_s              &  0.105 \ (0.025) &    0.107   &   0.01       \\
\;\;\;m_d/m_s          &  0.0521 \ (0.0067) &   0.0638    &   {\bf  3.09}       \\
\;\;\;Q^{-2} \times 10^3  &  1.934 \ (0.334)  &  1.815    &   0.12         \\
\;\;\;M_{\tau}         &  1.777 \ (0.1 \%)  &   1.777   &    < 0.01       \\
\;\;\;M_{\mu}          & 0.10566  \ (0.1 \%) &   0.10566  &  < 0.01      \\
\;\;\;M_e \times 10^3      &  0.511 \ (0.1 \%)&   0.511   &  < 0.01 \\
 \;\;\;V_{us}         &  0.22 \ (0.0026) &   0.2193  &   0.06      \\
\;\;\;V_{cb}         & 0.0413 \ (0.0015) &   0.0410   &   0.03         \\
\;\;\;V_{ub}    & 0.00367 \ (0.00047)  &  0.00316  &   {\bf 1.15}              \\
\;\;\;V_{td} &  0.0082 \ (0.00082) &  0.00824   &     < 0.01   \\
\;\;\;\epsilon_K          &  0.00228 \ (0.000228) &   0.00234  &    0.08       \\
\;\;\;\sin(2\beta) &      0.687 \ (0.064) &  0.6435 & 0.46  \\
\hline
\;\;\;\Delta m^2_{31} \times 10^3  &    2.3 \ (0.6) &  2.382 &  0.01  \\
\;\;\;\Delta m^2_{21} \times 10^5   &   7.9 \ (0.6) &  7.880   &  < 0.01  \\
\;\;\;\sin^2 \theta_{12} & 0.295 \ (0.045) &   0.289  &    0.01  \\
\;\;\;\sin^2 \theta_{23}  &  0.51 \ (0.13) &  0.532  &   0.03  \\
\hline
  \multicolumn{3}{|l}{{\rm TOTAL}\;\;\;\; \chi^2}  & {\bf 7.65}  \\
\hline
\end{array}
$$
\end{table}
\protect
\begin{table}
\caption[8]{ {\bf The fit for for fermion masses and mixing angles
at one particular point in SUSY parameter space defined by $m_{16} =
5$ TeV, $\mu = 300$ GeV and $M_{1/2} = 200$ GeV. } \\
   \mbox{Initial parameters: }\ \ \
\ \ (1/$\alpha_G, \, M_G, \, \epsilon_3$) = ($24.90,
\, 3.29 \times 10^{16}$ GeV, $\, -3.45$ \%), \makebox[1.8em]{ }\\
($\lambda, \, \lambda \epsilon, \, \sigma, \, \lambda \tilde \epsilon, \, \rho, \,
\lambda \epsilon', \, \lambda \epsilon \xi$) = ($ 0.63, \, 0.030, \, 0.77, \, 0.0070, \,
-0.054, \, -0.0022, \,
0.0035  $),\\
($\Phi_\sigma, \, \Phi_{\tilde \epsilon}, \, \Phi_\rho, \,
\Phi_\xi$) =  ($0.643, \, 0.410, \, 0.692, \, 3.618$) rad,
\makebox[6.6em]{ }\\
($m_{16}, \, M_{1/2}, \, A_0, \, \mu(M_Z)$) = ($5000,\, 200, \,
-9918.9, \,
300$) GeV,\\
($(m_{H_d}/m_{16})^2, \, (m_{H_u}/m_{16})^2, \, \tan\beta$) = ($1.91,  \, 1.62, \, 50.53$) \\
($M_{R_3}, \, M_{R_2}, \, M_{R_1}$) = ($6.1 \times 10^{13}$ GeV, $\,
-8.6 \times 10^{11} $ GeV, $\, 9.6 \times
10^{9} $ GeV) \\ 
} \label{t:fit}
$$
\begin{array}{|l|c|l|r|}
\hline
{\rm Observable}  &{\rm Data} \ (\sigma) & {\rm Theory} &  {\rm Pull} \\
{\rm (masses \ in \ GeV)}   &  &   &  \\
\hline
\;\;\;G_{\mu} \times 10^5   &  1.16637 \ (0.1 \%) & 1.16638  & < 0.01    \\
\;\;\;\alpha_{EM}^{-1} &  137.036 \ (0.1 \%) & 137.036  &  < 0.01        \\
\;\;\;\alpha_s(M_Z)    &  0.1187 \ (0.002) &   0.1178 &  0.18       \\
\hline
\;\;\;M_t              &  172.7 \ (2.9)   &   173.64  &       0.10 \\
\;\;\;m_b(M_b)          &    4.25 \ (0.25) &  4.29    &   0.03               \\
\;\;\;M_b - M_c        &    3.4 \ (0.2) &   3.49  &  0.23              \\
\;\;\;m_c(m_c)  &   1.2 \ (0.2)   &   1.08    &      0.34      \\
\;\;\;m_s              &  0.105 \ (0.025) &    0.113   &   0.12       \\
\;\;\;m_d/m_s          &  0.0521 \ (0.0067) &   0.0629    &   {\bf  2.60}       \\
\;\;\;Q^{-2} \times 10^3  &  1.934 \ (0.334)  &  1.824    &   0.10         \\
\;\;\;M_{\tau}         &  1.777 \ (0.1 \%)  &   1.777   &    < 0.01       \\
\;\;\;M_{\mu}          & 0.10566  \ (0.1 \%) &   0.10566  &  < 0.01      \\
\;\;\;M_e \times 10^3      &  0.511 \ (0.1 \%)&   0.511   &  < 0.01 \\
 \;\;\;V_{us}         &  0.22 \ (0.0026) &   0.2195  &   0.03      \\
\;\;\;V_{cb}         & 0.0413 \ (0.0015) &   0.0413   &    < 0.01         \\
\;\;\;V_{ub}    & 0.00367 \ (0.00047)  &  0.00325  &   0.77              \\
\;\;\;V_{td} &  0.0082 \ (0.00082) &  0.00812   &     < 0.01   \\
\;\;\;\epsilon_K          &  0.00228 \ (0.000228) &   0.00231  &    < 0.01       \\
\;\;\;\sin(2\beta) &      0.687 \ (0.064) &  0.6511 & 0.31  \\
\hline
\;\;\;\Delta m^2_{31} \times 10^3  &    2.3 \ (0.6) &  2.408 &  0.03  \\
\;\;\;\Delta m^2_{21} \times 10^5   &   7.9 \ (0.6) &  7.880   &  < 0.01  \\
\;\;\;\sin^2 \theta_{12} & 0.295 \ (0.045) &   0.288  &    0.01  \\
\;\;\;\sin^2 \theta_{23}  &  0.51 \ (0.13) &  0.537  &   0.04  \\
\hline
  \multicolumn{3}{|l}{{\rm TOTAL}\;\;\;\; \chi^2}  & {\bf 4.99}  \\
\hline
\end{array}
$$
\end{table}

\begin{table}
\caption[8]{Predictions for SUSY and Higgs spectra for the fit given
in Tables~\ref{t:fit4tev} and \ref{t:fit} in units of GeV.}
\label{t:susy_and_higgs}
$$
\begin{array}{|l|c|c|}
\hline
{\rm Particle}  & m_{16}=5 \ {\rm TeV} &m_{16}=4 \ {\rm TeV} \\
\hline
\;\;\; h              & 120  & 124\\
\;\;\; H              & 699  & 699\\
\;\;\; A^0            & 700  & 699\\
\;\;\; H^+            & 700  & 701\\
\;\;\; \chi^0_1       & 81   & 80\\
\;\;\; \chi^0_2       & 151  & 150\\
\;\;\; \chi^+_1       & 151  & 150\\
\;\;\; \tilde g       & 605  & 597\\
\;\;\; \tilde t_1     & 498  & 496\\
\;\;\; \tilde b_1     & 902  & 846\\
\;\;\; \tilde \tau_1  & 1686 & 1421\\
\hline
\end{array}
$$
\end{table}
\begin{table}
\caption[8]{Predictions for neutrino masses, $\sin^2\theta_{13}$ and
CP violation in the lepton sector for the fit given in
Tables~\ref{t:fit4tev} and \ref{t:fit}.}
\label{t:neutrino_predictions}
$$
\begin{array}{|l|c|c|}
\hline
  & m_{16}=5 \ {\rm TeV} &m_{16}=4 \ {\rm TeV} \\
\hline
\;\;\; m_{\nu_3} \ {\rm (eV)} \;\; &   0.0492  &0.0489  \\
\;\;\; m_{\nu_2} \ {\rm (eV)} \;\; &   0.0097  &0.0096  \\
\;\;\; m_{\nu_1} \ {\rm (eV)} \;\; &   0.0041  &0.0036\\
\;\;\;\sin^2 \theta_{13}\;\;       &  0.0025   &0.0037    \\
\;\;\; J                            &     0.0013&0.013   \\
\;\;\; \sin \delta      \;\;        &     0.119&0.996   \\
\;\;\; \alpha_1 \ {\rm (rad)}      \;\;        & -2.974 &-1.771 \\
\;\;\; \alpha_2  \ {\rm (rad)}     \;\;        & 0.136  &1.315\\
\;\;\;  \left< m_{\beta \beta} \right>  \ {\rm(eV)}    \;\;        & 0.00010 &0.00038 \\
\;\;\;   m_{\nu_e}^{eff}   \ {\rm (eV)}    \;\;        & 0.0067  &0.0067\\
\;\;\; \epsilon_1       \;\;        & 0.92 \times 10^{-7}  &1.61\times 10^{-7}\\
\hline
\end{array}
$$
\end{table}
\begin{table}
\caption[8]{Predictions for branching ratios for lepton flavor
violating processes and the electric dipole moment of leptons for
the fit given in Tables~\ref{t:fit4tev} and \ref{t:fit}.}
\label{t:lfv_and_edm}
$$
\begin{array}{|l|c|c|}
\hline
  & m_{16}=5 \ {\rm TeV} &m_{16}=4 \ {\rm TeV} \\
\hline
\;\;\; Br(\mu\to e\gamma)   &4.69\times 10^{-15}& 1.40\times 10^{-14}\\
\;\;\; Br(\tau\to e\gamma)  &1.26\times 10^{-12}& 2.40\times 10^{-12}\\
\;\;\; Br(\tau\to \mu\gamma)&6.13\times 10^{-11}& 1.22\times 10^{-10}\\
\;\;\; d_e \ [e \cdot cm]    &9.15\times 10^{-30}  & 2.43\times 10^{-29}\\
\;\;\; d_{\mu} \ [e \cdot cm]&-3.58\times 10^{-28}   & -7.93\times 10^{-28}\\
\;\;\; d_{\tau}\ [e \cdot cm]&-1.54\times 10^{-27}  & -2.64\times 10^{-27}\\
\hline
\end{array}
$$
\end{table}

\section{Summary and Conclusions}

In this paper we have performed a global $\chi^2$ analysis on a
well-motivated, phenomenologically acceptable minimal SO(10) SUSY
GUT with a $D_3$ family symmetry.   The most stringent constraint
comes from assuming Yukawa coupling unification for the third family
of quarks and leptons.  The $\chi^2$ contours as functions of $\mu$
and $M_{1/2}$ for $m_{16} =$ 4 and 5 TeV are given in Figs. 1 and 2.
We find acceptable solutions with $\chi^2 < 8$ (6) for $m_{16} =$ 4
(5) TeV, respectively.     We find the light Higgs mass, found using
FeynHiggs,  has an upper bound given by $m_h \leq 127$ GeV (see
Figs. \ref{fig:mh4tev} and \ref{fig:mh}). The additional predictions
for the SUSY spectrum, and neutrino masses and mixing angles at two
particular points in SUSY parameter space are given in Tables
\ref{t:susy_and_higgs} and \ref{t:neutrino_predictions}.

In addition to the global $\chi^2$ analysis, we focused on obtaining
the rates for several lepton flavor violating processes and also for
the charged lepton electric dipole moments. We calculated the
branching ratios for the lepton flavor violating processes $Br(\mu
\rightarrow e \gamma)$ (Figs. \ref{fig:muegamma4tev} and
\ref{fig:muegamma}) and $\tau \rightarrow e \gamma$ and $\tau
\rightarrow \mu \gamma$ (Fig. \ref{fig:tau2lgamma}).  There is only
a narrow region in the upper right hand corner (Fig.
\ref{fig:muegamma4tev}) with $\chi^2 \leq 8$ which is within the
sensitivity of the MEG experiment.   We have also evaluated the
electric dipole moments of the electron (Figs. \ref{fig:edmE4tev}
and \ref{fig:edmE}) and muon and tau (Fig. \ref{fig:edmMT}). In all
cases the results are below the present experimental bounds, and for
$d_\mu$ the result is below the projected sensitivity of future
experiments \cite{Farley:2003wt}. However, in both cases, the entire
region is within the projected sensitivity of future experiments for
$d_e$ \cite{Kawall:2004nv}.  The results for the lepton flavor
violating processes and electric dipole moments at the same two
particular points in SUSY parameter space are given in Table
\ref{t:lfv_and_edm}.

\vspace{.5in} \noindent {\bf Acknowledgements} 
We would like to thank T. Bla\v{z}ek for the use of his code. We also thank
M. Albrecht, A. Buras and D. Straub for pointing out typos in captions of tables 2 and 3 in previous 
versions of this paper to us and for comparing some numerical results.
M.H. and S.R. are
partially supported by DOE grant DOE/ER/01545-868.  R.D. is 
supported by the U.S. Department of Energy, grant DE-FG02-90ER40542.

\end{document}